\documentclass[preprints,article,accept,moreauthors,pdflatex]{Definitions/mdpi} 

\firstpage{1} 
\pubvolume{xx}
\issuenum{1}
\articlenumber{5}
\pubyear{2019}
\copyrightyear{2019}
\history{}
\pdfoutput=1

\usepackage[figuresright]{rotating}
\usepackage{amsmath}
\usepackage{lastpage}
\usepackage{array}
\usepackage{subcaption}
\usepackage[utf8]{inputenc}
\usepackage{array}
\usepackage{makecell}
\usepackage{textcomp}
\usepackage[T1]{fontenc}
\usepackage{color}
\usepackage{nomencl}
\usepackage{etoolbox}
\usepackage{amssymb}
\usepackage{mathtools}
\usepackage{adjustbox}
\usepackage{placeins}
\renewcommand\nomgroup[1]{%
  \item[\bfseries
  \ifstrequal{#1}{A}{Parameters}{%
  \ifstrequal{#1}{B}{Subscripts}{%
  \ifstrequal{#1}{C}{Other Symbols}{}}}%
]}
\newcommand{\nomunit}[1]{%
\renewcommand{\nomentryend}{\hspace*{\fill}#1}}
\renewcommand{\d}{\text{d}}
\newcommand{\dt}{\text{dt}}
\usepackage{xstring}
\usepackage[nolist]{acronym}
\usepackage[american]{circuitikz}
\ctikzset{bipoles/resistor/width=.7}
\ctikzset{bipoles/resistor/height=.36}

\newcommand{\n}[1]{_{\mathrm{#1}}}

\newcommand{\gA}{g\n{A}}
\newcommand{\Cw}{C\n{e}}

\newcommand{\Ci}{C\n{i}}
\newcommand{\Cf}{C\n{f}}

\newcommand{\Ps}{\Phi\n{s}}
\newcommand{\Ph}{\Phi\n{h}}
\newcommand{\Rwa}{R\n{e,a}}

\newcommand{\Riw}{R\n{i,e}}

\newcommand{\Rfi}{R\n{f,i}}

\newcommand{\Ta}{T\n{a}}
\newcommand{\Tw}{T\n{e}}

\newcommand{\Ti}{T\n{i}}
\newcommand{\Tf}{T\n{f}}

\begin{acronym}
\acro{CHP}{Combined Heat and Power}
\acro{MPC}{Model Predictive Control}
\end{acronym}
\Title{Control of heat pumps with CO$_2$ emission intensity forecasts}

\Author{Kenneth Leerbeck $^{1,\dagger, \text{§}}$*, Peder Bacher $^{2, \text{§}}$\orcidA{}, Rune Gr{\o}nborg Junker $^{3, \text{§}}$\orcidB{}, Anna~Tveit $^{4,§}$, Olivier~Corradi $^{5,¶}$ and Henrik Madsen $^{6,\text{§}, ‖}$\orcidC{} }

\AuthorNames{Kenneth Leerbeck, Peder Bacher, Rune Gr{\o}nborg Junker, Anna~Tveit and Henrik Madsen}

\address{%
$^{1}$ \quad Affiliation 1; kenle@dtu.dk\\
$^{2}$ \quad Affiliation 2; pbac@dtu.dk\\
$^{3}$ \quad Affiliation 3; rung@dtu.dk\\
$^{4}$ \quad Affiliation 4; annatveit@hotmail.com\\
$^{5}$ \quad Affiliation 5; olivier.corradi@tmrow.com\\
$^{6}$ \quad Affiliation 6; hmad@dtu.dk}

\corres{Correspondence: Affiliation 1; Tel.: +45-61427386 (F.L.)}

\firstnote{Current address: Affiliation 1} 
\thirdnote{Technical University of Denmark, DTU} 
\fourthnote{Technical University of Norway, NTU} 
\fifthnote{Tomorrow IVS}
\abstract{An optimized heat pump control for building heating was developed for minimizing CO$_2$ emissions from related electrical power generation. The control is using weather and CO$_2$ emission forecasts as input to a Model Predictive Control (MPC) - a multivariate control algorithm using a dynamic process model, constraints and a cost function to be minimized. In a simulation study the control was applied using weather and power grid conditions during a full year period in 2017-2018 for the power bidding zone DK2 (East, Denmark).\\
Two scenarios were studied; one with a family house and one with an office building. The buildings were dimensioned on the basis of standards and building codes. The main results are measured as the CO$_2$ emission savings relative to a classical thermostatic control. Note that this only measures the gain achieved using the MPC control, i.e.\ the energy flexibility, not the absolute savings. The results show that around 16\% savings could have been achieved during the period in well insulated new buildings with floor heating.
\\
Further, a sensitivity analysis was carried out to evaluate the effect of various building properties, e.g.\ level of insulation and thermal capacity. Danish building codes from 1977 and forward was used as benchmarks for insulation levels. It was shown that both insulation and thermal mass influence the achievable flexibility savings, especially for floor heating. Buildings that comply with building codes later than 1979 could provide flexibility emission savings of around 10\%, while buildings that comply with earlier codes provided savings in the range of 0-5\% depending on the heating system and thermal mass.}

\begin{document}
% \biboptions{numbers,sort&compress}
%%%%%%%%%%%%%%%%%%%%%%%%%%%%%%%%%%%%%%%%%%
%%%%%%%%%%%%%%%%%%%%%%%%%%%%%%%%%%%%%%%%%%

%The order of the section titles is: Introduction, Materials and Methods, Results, Discussion, Conclusions for these journals: aerospace,algorithms,antibodies,antioxidants,atmosphere,axioms,biomedicines,carbon,crystals,designs,diagnostics,environments,fermentation,fluids,forests,fractalfract,informatics,information,inventions,jfmk,jrfm,lubricants,neonatalscreening,neuroglia,particles,pharmaceutics,polymers,processes,technologies,viruses,vision
\section{Introduction}
\label{Section:Introduction}
\small

Energy flexibility on the electricity market is a high focus area in modern energy policies scoping in on storage (e.g. batteries, fuel cells, hydro reservoirs, thermal) and flexible demand (e.g. heat pumps, electric cars), \cite{IEA_sol}. The aim is to decrease CO$_2$ emissions by meeting the fluctuating proportion of renewable sources (eg. solar, wind) vs. nonrenewable sources (eg. coal, gas, nuclear). Ideally, in the future, electricity users (the demand) will respond to the renewable power generation levels in attempt to minimize emissions - in a 100\% renewable scenario storage and flexibility is a must for operating the power system \cite{HUBER}. 

Therefore, methods for identifying the flexibility potential in various applications are developed. In \cite{Groenborg}, the energy flexibility potential in buildings is identified by the so called \textit{Flexibility Index}, which is the energy cost, from a penalty-aware control, relative to a penalty-ignorant control. The penalty could be e.g. a CO$_2$ or price signal. The present paper investigates the energy flexibility potential in buildings with a focus on heat pumps.

Heat pumps have different sizes and applications, from small single building- to large heat pumps for district heating. The scope of the present study is limited to investigate the increasing potential in single building heat pumps - which has been almost four-fold from 2011 to 2019 while the number of oil fired boilers have decreased by roughly one third in the same period, \cite{dst}. Many oil fired boilers are replaced with heat pumps - due to both economic and environmental benefits and political pressure (bans of oil fired boilers in certain districts for new buildings, \cite{mininsulation}.) The control of the heating, however, are often simple thermostatic controls. This often results in heating when electricity demand is high (e.g. afternoon and evening peaks), leading to increased system stress, resulting in increased fossil fuel consumption. It is therefore an opportunity to shift the demand away from peak hours using the heat storing potential of the buildings.

\begin{figure}[t]
\centering
  \includegraphics[width=\linewidth/2]{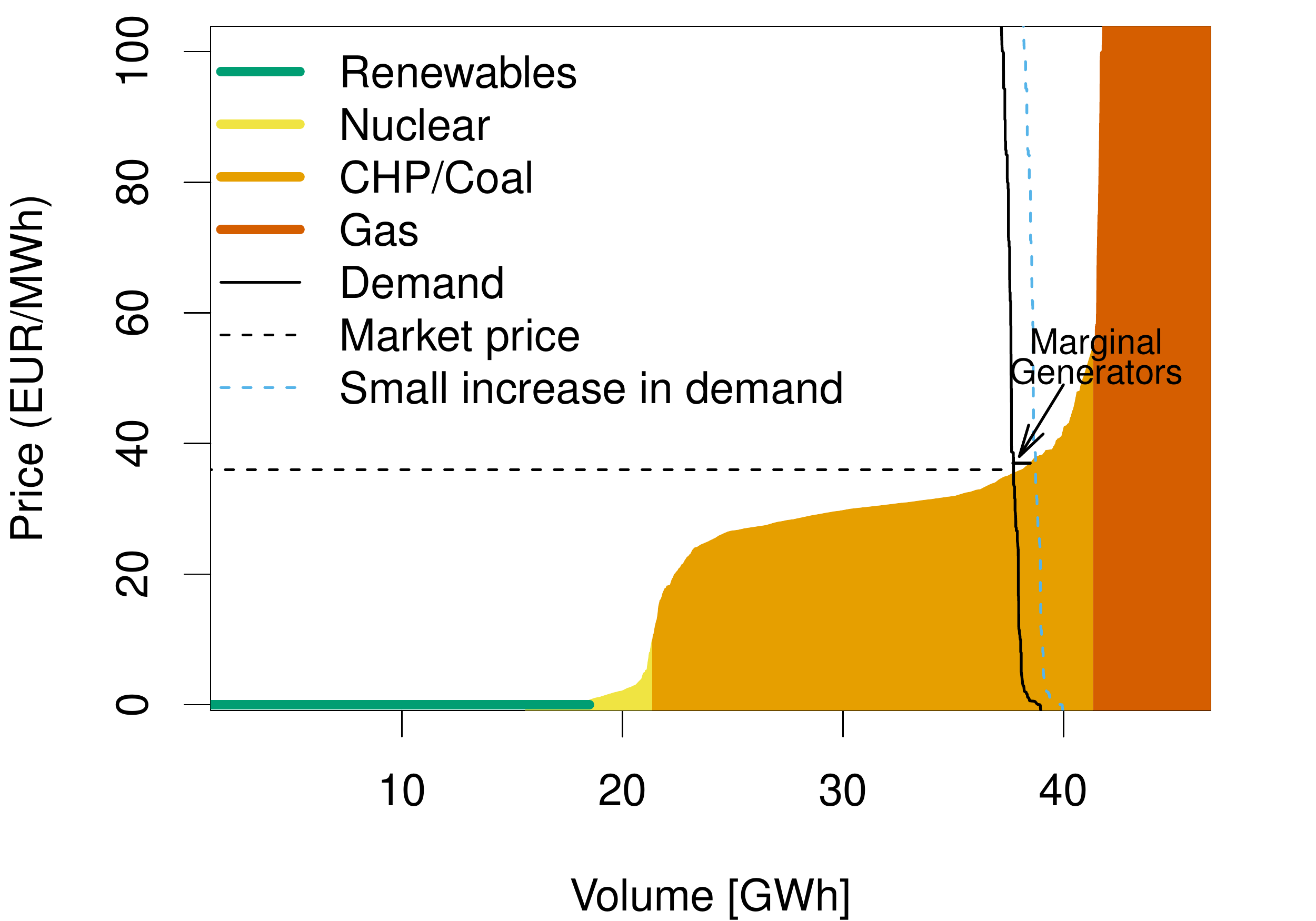}
  \caption{The merit order illustrated with a supply and demand curve example. The x-axis is the accumulated generators in the power system and the y-axis is their corresponding costs. The highest generator in the merit order is the one crossing with the demand curve - a coal \ac{CHP} plant in this example. The average emissions are a weighted average from all activated generators. The\textbf{ marginal generator} is the generator that will be activated by moving the demand line slightly to the right (dashed blue line). Data source: Nord Pool AS.}
  \label{fig:merit}
\end{figure}

In a power system the generator which is responding to small changes in demand (e.g. start-up of a heat pump) is called the marginal generator. A good estimate of the marginal generator is achieved by using price signals, see Figure~\ref{fig:merit} - the merit order illustrated with a supply/demand curve; the x-axis has the accumulated supply generators and the y-axis is the corresponding price. A small increase in demand (dashed blue line) illustrates the marginal generator - in this case a coal fired \ac{CHP} plant.

Due to both grid stability, economic and environmental benefits, day-ahead spot price-based control strategies have been proposed in recent papers \cite{Halvgaard2012, Corradi2,JochClaus}, using occupancy mode detection and rule-based price control and \ac{MPC} (a multivariate predictive control algorithm using a dynamic process model, constraints and a cost function to be minimized).
In \cite{Halvgaard2012}, \ac{MPC} is used with varying electricity prices to minimize the cost for operating a heat pump connected to a storage unit and a floor heating system. The control only heats at night, where the prices are low, and it is assumed that the heat pump and storage are large enough to accumulate enough heat for the whole day. Cost savings of between 25\% to 30\% are obtained. \ac{MPC} is a well known concept in building automation control literature,  \cite{Olde,Olde2,Killian,Linde,Chen}, and proven to be promising w.r.t. minimizing costs, but a broad practical implementation still has various challenges discussed in \cite{Killian2}. 

In \cite{Olde2}, the importance of occupancy information is highlighted and evaluated on a daily basis. However, a higher resolution is needed to incorporate variations throughout the day (e.g. when people are at work).
In the study \cite{Corradi2} occupancy modes are used together with price signals to control a heat pump. The occupancy modes were developed in The Olympic Peninsula project \cite{Hammerstrom} and describe \textit{work, night} and \textit{home} mode, each with a corresponding set point and price sensitivity. The study showed a significant level of load shifting, leveling out the normal peaks in the daily demand curve. A self-learning controller was applied and adapts easily to changing consumer habits.

There is a problem with spot prices though, known as the merit order emission dilemma, as illustrated in \cite{Wolfgang} for the German-Austrian power market: The price for coal is low but the emissions are high. A price-based control, therefore, only leads to a decrease in emissions if there is surplus of renewable energy (more renewable energy than needed) - otherwise coal is favored, and it is therefore encouraged to use CO$_2$ emission signals instead. 

For CO$_2$ emissions, two distinct measures are used: average and marginal emission intensities, both with the units $\left(\frac{\text{kgCO}_2\text{-eq}}{\text{MWh}}\right)$. Average emissions correspond to the overall, e.g. region-wide, electricity production including net imports. The marginal reflects the emissions of the marginal generator. The concepts are compared in \cite{Anika} and the importance of distinguishing between the two is highlighted due to their very opposing patterns. It is emphasized that the marginal emission is the most optimal signal to use for control.

In \cite{JochClaus}, the average CO$_2$ emission intensity and price signals are used in heat pump control of residential buildings in Norway (known for low emissions due to large amount of hydro power) with Predictive Rule-Based Control (uses predefined thresholds to give information about when the emissions are low). 
it is concluded that with price-based control, the overall CO$_2$ emissions have actually increased (evaluated using the average emissions).
It is argued to result from the load being shifted to the night time, where cheap carbon-intensive electricity is imported from the continental European power grid. This is either a great example of the merit order dilemma or a result that may have been different if marginal emissions had been used.

A recent study, \cite{KL}, investigates marginal emissions and uses estimates provided by Tomorrow \footnote{\url{www.tmrow.com}} to develop a 24-hour forecast using a machine learning approach on historical data. The CO$_2$ estimates are calculated with the empirical approach developed in \cite{Corradi} using historical data from European bidding zones. The chain of imports (the so-called flow tracing, originally introduced in \cite{Bialek,Kirschen}) is followed to assess the impact of a specific generator or load on the power system. This is a large scale solution using data from the majority of bidding zones around the world.

In the present study \ac{MPC} is used for control for heating a building. This allows using knowledge of future indoor climate states, CO$_2$ emissions and weather conditions to schedule heat pump operation. It is a linear approach, which has its limits and requires simplifications, investigated and discussed in \cite{Clara2011}. The simplifications include neglecting the effect on the efficiency from factors e.g.\ frequency variations in the compressor (a main component in a heat pump) and temperature variations. The paper concludes that neglecting these factors can lead to significant errors. The frequency variations are, however, neglected in this study. From \cite{Clara2011}, the frequency is noticed to be the least important factor and is specifically justified when using varying electricity prices, because the heat pump mostly operates at nominal speed to maximize the heat output when prices are low. The impact from the outdoor temperature is accounted for - this is important because it means the efficiency is lower during the night, where also the emissions are low.

In order to model the heat dynamics in the building a lumped dynamic process model is applied \cite{Springer2}. A tricky part is to determine the values of the parameters appropriately, e.g.\ insulation level and heat capacity: if the right type of measured data is available the parameters can be estimated, \cite{Madsen1995}, or they can be calculated according to physics. In the present study physics are used and a sensitivity analysis is carried out to map the impact of parameters on the CO$_2$ savings potential. Such a sensitivity analysis is lacking in the literature. In some papers transparency is lost, since the impact of the parameters is not elucidated, thus increasing the risk of biased results. This paper addresses both of these issues by using historic danish building codes from 1977 and later to describe the insulation thickness as a parameter along with the heat pump size and thermal capacity of floor in two hypothetical buildings: a family house and an office building. Further, the impact of using forecasts is assessed by comparing the savings achieved with known future weather (perfect forecasts) vs. real forecasts.

It is noted that the emission saving potential using an \ac{MPC}, i.e.\ flexible demand, is measured as CO$_2$ emission savings relative to a classical thermostatic control, i.e.\ non-flexible demand. Hence, the results express only the potential of energy flexibility, not the absolute emission savings.

% The approaches are continuously improved to better fit the real world. There are three main concepts to distinguish between; black box models - basing building parameters purely on material properties and physical equations; white box - basing building parameters purely on data points; grey box where data points and physical equations are used in combination to create the most accurate model. A study conducted by Henrik Madsen and Jan Holst in 1995 \cite{Madsen1995} is one of the earliest studies that succeeded with grey box modelling and their approach is still used today in the \textsf{R}-package CTSM-R (Continous Time Series Modelling in \textsf{R}). 
 
In Sec. \ref{sec:data}, the weather data and marginal CO$_{2}$ emissions are presented. The dynamic process model is presented in Sec. \ref{sec:Model} as an RC-diagram together with the \ac{MPC} which is written up as a linear programming formulation. The efficiency of the heat pump is modelled as a temperature  dependent variable, but neglects the compressor frequency. In Sec. \ref{Section:Buildings} the building codes, temperature settings and model parameters are discussed. The results are presented in Sec. \ref{section:Results} as graphs showing the CO$_2$ emission reductions vs selected parameters - e.g. heat pump size, concrete and building regulation year. 

%Our aim is to demonstrate that the forecasts developed in \cite{KL} can be used to reduce the CO$_2$ emissions related to operating a flexible application e.g. heat pump. Furthermore, to map the range of buildings that could benefit from the MPC.

\printnomenclature

\section{Data}\label{sec:data}

\begin{figure}[!tbp]
  \centering
  \begin{minipage}[b]{0.49\textwidth}
  \includegraphics[width=\linewidth]{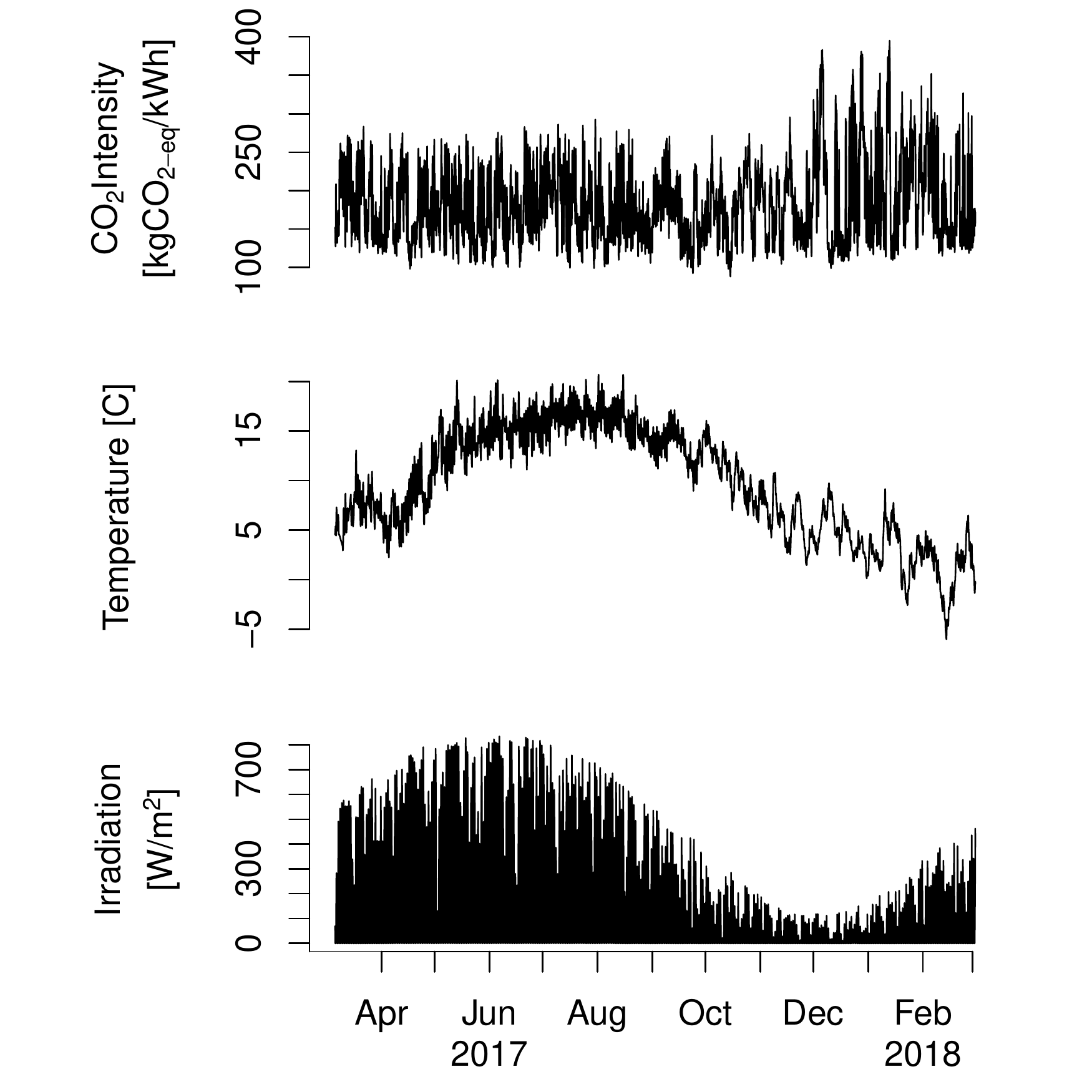}
  \caption{Marginal CO$_2$ emissions (real-time), Temperature and solar irradiation (forecasts) plotted for the evaluated period.}
  \label{fig:weather}
  \end{minipage}
  \hfill
  \begin{minipage}[b]{0.49\textwidth}
  \includegraphics[width=\linewidth]{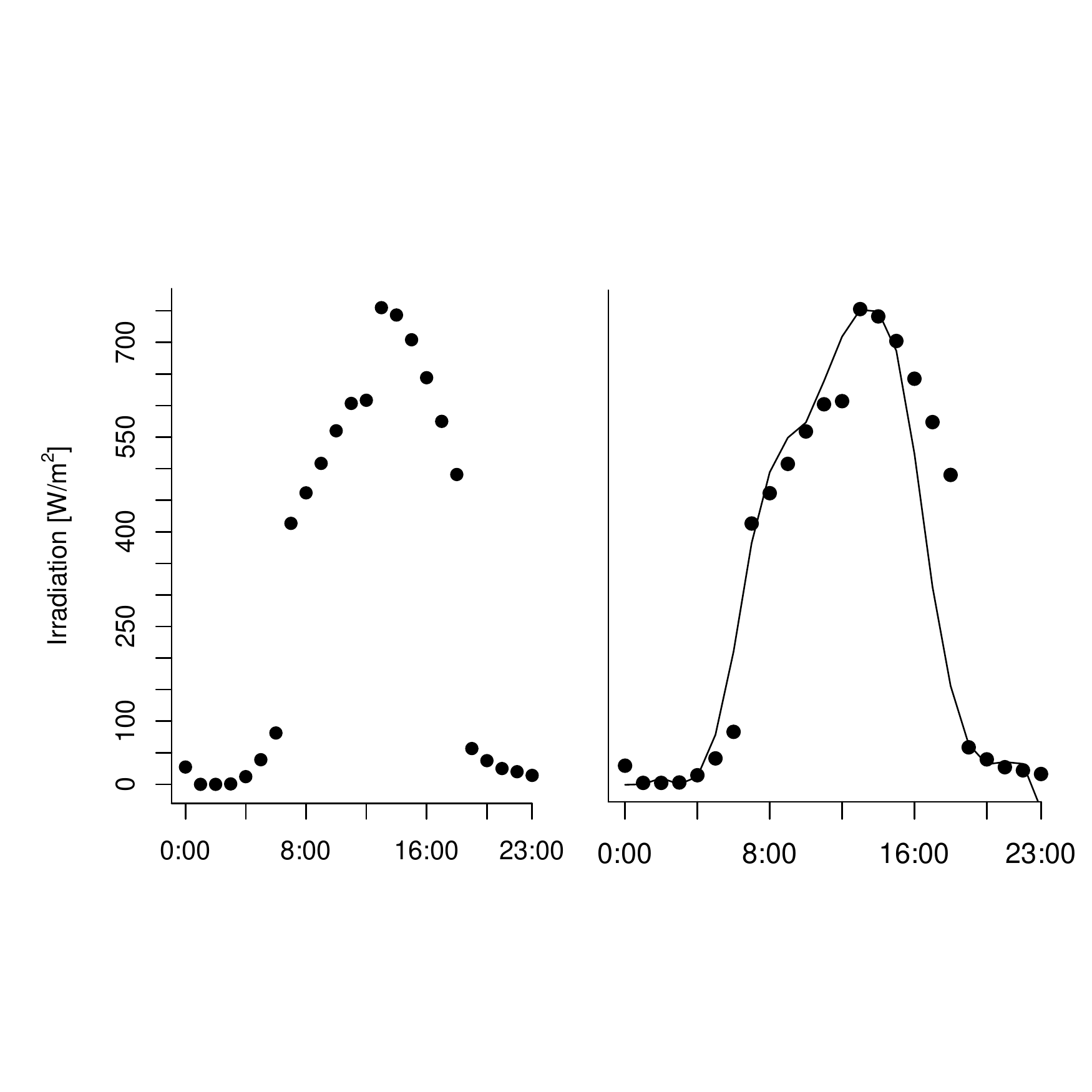}
  \caption{i) Solar irradiation 6 hour horizon forecasts (left) for the 21st of June 2017 (updated at 2am,8am,14pm and 20pm). ii) Model of the real-time irradiation derived from a kernel smoothing approach on the updated forecasts (right).}
  \label{fig:solar}
  \end{minipage}
\end{figure}

Data provided by Tomorrow \footnote{\url{www.tmrow.com}} is used in the study. It comprises the marginal CO$_2$ emission data and weather forecasts (temperature and solar irradiation) to model the building thermodynamics and heat pump planning - see Figure~\ref{fig:weather}. The emissions show close to none seasonality except from the winter, where the intensity peaks. Both the temperature and solar radiation are highly seasonal with hot climate and high solar radiation in the summer - reversed in the winter.

Only the CO$_2$ emissions are provided in both real-time values and forecasts. The real-time weather conditions are, however, also very important for the model to describe the actual building thermodynamics. The solar irradiation forecast is plotted  in Figure~\ref{fig:solar} (left plot) for the 21st of June. Every sixth hour, at 2am,8am,14pm and 20pm, a new forecast is provided. The gaps between the forecasts are significant and illustrates the inaccuracy of the prediction for long horizons. 

For modelling purposes, the real-time weather conditions are modelled from the forecasts. Assuming horizons for $h=1$ are the most accurate forecasts, a kernel smoothing process using splines and a weight for short horizon favouritism is applied - see Appendix \ref{Appendix:kernel} for a description of the approach. The result for solar radiation is the smoothing curve seen in Figure~\ref{fig:solar}, right plot. The temperature forecasts did not show any significant gaps, suggesting all horizons, 1-6, are more correct models of the real-time condition than the solar radiation.

\section{Model}
\label{sec:Model}
\small
%The \textsf{R} package CTSM-R provides the framework for the stochastic model that is used for this study.

The model is a state space model, derived from thermodynamic state equations describing the heat dynamics in the building as a lumped dynamic process model, \cite{TSA}. The model parameters are defined based on the building composition and structure. However, if the right measurement data is available the model parameters could be estimated as in \cite{Bacher2011}, thus it would be easy to use the applied control setup in existing buildings, without the need for information about the building composition.

% For more information of CTSM-R see the CTSM-R manual \textbf{\cite{ctsmr}}.

\subsection{Assumptions}
\label{sec:assumptions}
%A lumped system analysis is used to reduce the thermal system to a number of "lumps", where it is assumed that the temperature differences inside each lump are negligible. This approximation is useful to simplify otherwise complex differential heat equations.

It is assumed that the building is just one big room with a flat roof. Thereto the following assumptions have been made; one uniform air temperature; no ventilation; no influence from humidity of the air; no influence from wind.
The heat pump is assumed to be static because its dynamics are much faster than those of the building. Heat pumps require power to move heat from a cold space to a hot space using a refrigerant (it extracts heat from a cold space through an evaporator and delivers that heat to the hot space through a condenser). The heat pump efficiency is described using the COP-factor (Coefficient Of Performance). The maximum efficiency is modelled as the Carnot Efficiency\cite{McGraw} by
\begin{align}\label{eq:COP}
    \text{COP}_{\text{Carnot}} = \left(1 - \frac{T\n{cold}}{ T\n{hot}}\right)^{-1},
\end{align}
where $T\n{hot}$ represents the condensation temperature, which is the temperature of the water flowing in the heating system (fluctuating in reality, but simplified to a constant $=$ 40\textdegree{}C). $T_{cold}$ is the ambient evaporation temperature.
However, the COP-factor is smaller in reality and therefore it is multiplied by another efficiency $\eta$, which can be assumed to be between 50\% and 70\%, \cite{indu}. To be conservative, it has been set to 50\% in the calculations. Therefore, the COP factor is expressed as
\begin{align}\label{eq:COPCOP}
    f\n{cop}(T\n{hot},T\n{cold}) = \eta \cdot \text{COP}_{\text{Carnot}}.
\end{align}

\subsection{State Space Equations}
The state space model is defined by dividing the building into three sections; '\textit{Floor}', '\textit{Interior}', and '\textit{Inner envelope}'. \textit{Inner} refers to the part of the walls and roof that is on the inside of the insulation.

The goal is to determine the temperature dynamics in the three sections. Thermodynamics can very well be explained in the same way as electric circuits, which will provide a nice analytic approach. The model is shown as the commonly used RC-diagram, \cite{Madsen1995}, in Figure \ref{fig:model_diagram}.
\begin{figure}[ht]
  \centering
  %% Variable for scaling the tikz figure
  \newcommand{\scaleTikz}{0.45}
  %% Control text size
  \large
  %% Scaling the nodes
\tikzset{every node/.style={scale=\scaleTikz}}
%% The tikz
\begin{circuitikz}[xscale=\scaleTikz, yscale=\scaleTikz]
  %% Define variables to control size
  \def\x{2.3}    % x distance between nodes
  \def\y{3.3}    % y position of the diagram
  \def\ylab{4} % y position of the labels
  %% Set the coordinates
  \coordinate (Floor) at (0,0);
  \coordinate (Heater) at ($(Floor)+(0.8*\x,0)$);
  \coordinate (SolarFloor) at ($(Heater)+(\x,0)$);
  \coordinate (Interior) at ($(SolarFloor)+(\x,0)$);
  \coordinate (SolarInterior) at ($(Interior)+(1.3*\x,0)$);
  %\coordinate (HeaterInterior) at ($(Interior)+(\x,0)$);
  \coordinate (WallLeft) at ($(SolarInterior)+(0.45*\x,0)$);
  \coordinate (Wall) at ($(WallLeft)+(0.8*\x,0)$);
  \coordinate (WallRight) at ($(Wall)+(0.8*\x,0)$);
  \coordinate (Ambient) at ($(WallRight)+(0.25*\x,0)$);
  %% Draw stuff
  \draw
  ($(3.15*\x,0)$) node[ground] {}
  % ($(6.15*\x,0)$) node[ground] {}

 % ($(Floor)+(0,4.5)$) -- ($(SolarFloor)+(0,4.5)$)
 % ($(Floor)+(0,6)$) -- ($(SolarFloor)+(0,6)$)
 
  %($(Heater)+(\xoffset,0)$) -- ($(Heater)+(\xoffset,1)$)
  %($(Heater)+(\xoffset,1)$) to[I, l=$\Ph$] %($(Heater)+(\xoffset,\y)$);
 
  (Floor) -- (Ambient) % The bottom line
  ($(Floor)+(0,\y)$) -- ($(SolarFloor)+(0,\y)$) % The top line
  ($(Interior)+(0,\y)$) -- ($(WallLeft)+(0,\y)$); % The top line

  % The top line

  %% The Floor
  \draw
  (Floor) to[C=$\Cf$, -*] ($(Floor)+(0,\y)$)
  {[anchor=south east] ($(Floor)+(0.2,\y)$) node {$\Tf$}};
  %% The Heater
  \def\xoffset{-0.25*\x}
  \draw
  ($(Heater)+(\xoffset,0)$) -- ($(Heater)+(\xoffset,1.2)$)
  ($(Heater)+(\xoffset,1.2)$) to[I, l_=$(1-\Psi\n{h})\Ph$] ($(Heater)+(\xoffset,\y)$);
  
  %% The Solar to floor
  \def\xoffset{0.1*\x}
  \draw
  ($(SolarFloor)+(\xoffset,0)$) to[I, l=$(1-\Psi\n{s}) \gA\Ps$]($(SolarFloor)+(\xoffset,\y-1.2)$)
  ($(SolarFloor)+(\xoffset,\y-1.2)$) -- ($(SolarFloor)+(\xoffset,\y)$);
  %% Resistance
  \draw
  ($(SolarFloor)+(0,\y)$) to [R, l=$\Rfi$] ($(Interior)+(0,\y)$);
  %% The Interior
  \draw
  (Interior) to[C=$\Ci$, -*] ($(Interior)+(0,\y)$)
  {[anchor=south east] ($(Interior)+(0.2,\y)$) node {$\Ti$}};
  %% The Solar to interior
  \def\xoffset{0.1*\x}
  \draw
  ($(Heater)+(\xoffset+5.5,0)$) -- ($(Heater)+(\xoffset+5.5,1.2)$)
  ($(Heater)+(\xoffset+5.5,1.2)$) to[I, l_=$\Psi\n{h}\Ph$] ($(Heater)+(\xoffset+5.5,\y)$);
  \def\xoffset{0*\x}
  \draw
  ($(SolarInterior)+(\xoffset+0.5,0)$) to[I, l=$\Psi\n{s}\gA\Ps$]($(SolarInterior)+(\xoffset+0.5,\y-1.2)$)
  ($(SolarInterior)+(\xoffset+0.5,\y-1.2)$) -- ($(SolarInterior)+(\xoffset+0.5,\y)$);
  %% The Wall
  \draw
  ($(WallLeft)+(0,\y)$) to [R, l=$\Riw$, -*] ($(Wall)+(0,\y)$)        
  ($(Wall)+(0,\y)$) to [R, l=$\Rwa$] ($(WallRight)+(0,\y)$)
  ($(WallRight)+(0,\y)$) -- ($(Ambient)+(0,\y)$)
  (Wall) to[C=$\Cw$, -] ($(Wall)+(0,\y)$)
  {[anchor=south east] ($(Wall)+(0.2,\y)$) node {$\Tw$}};
  %% The Ambient
  \draw
  [V, l=$\Ta$] ($(Ambient)+(0,\y)$) to ($(Ambient)+(0,0)$);
  %% The dashed lines
  \def\xoffset{0.22*\x}
  \draw
  [dashed] ($(SolarFloor)+(\xoffset*\x,-0.5)$) -- ($(SolarFloor)+(\xoffset*\x,\y-0.4)$)  
  [dashed] ($(SolarFloor)+(\xoffset*\x,\ylab+0.2)$) -- ($(SolarFloor)+(\xoffset*\x,\ylab+1)$)  
  [dashed] ($(WallLeft)+(0.15,-0.5)$) -- ($(WallLeft)+(0.15,\ylab+1)$)  
  [dashed] ($(Ambient)+(-0.3*\x,-0.5)$) -- ($(Ambient)+(-0.3*\x,\ylab+1)$);
  %% The labels
  \draw
  {[anchor=south] ($(Heater)+(0.2,\ylab)$) node{\Large Floor}}
  {[anchor=south] ($(Interior)+(0.7*\x,\ylab)$) node{\Large Interior}}
  {[anchor=south] ($(Wall)+(0,\ylab)$) node{\Large Envelope}}
  {[anchor=south] ($(Ambient)+(0.2*\x,\ylab)$) node{\Large Ambient}};

\end{circuitikz}
  \caption{RC-network diagram of the model (floor heating).}
  \label{fig:model_diagram}
\end{figure}
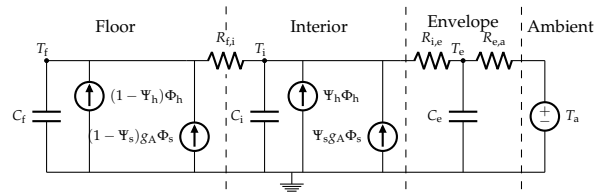
\begin{align}\label{eq:Tm}
    \frac{\d T^f_t}{\dt} &= \frac{1}{\Cf} \left( \frac{T^i_t-T^f_t}{\Rfi} + (1 - \Psi_s) \gA \Phi^s_t+ (1-\Psi_h) \Phi^h_t
    \right)\\
\label{eq:Ti} 
    \frac{\d T^i_t}{\dt} &= \frac{1}{\Ci} \left( \frac{T^f_t-T^i_t}{\Rfi} + \frac{T^w_t-T^i_t}{\Riw} +
    \Psi_s \gA \Phi^s_t+ \Phi^h_t\Psi_h \right)
   \\
\label{eq:Tw} 
    \frac{\d T^w_t}{\dt} &= \frac{1}{\Cw} \left( \frac{T^i_t-T^w_t}{\Riw} + \frac{T_a-T^w_t}{\Rwa}\right)
\end{align}
Refer to the nomenclature in Section \ref{Section:Introduction} for variable definitions. $\Psi_h$ is a logic variable integer (0,1) defining the heat system ($\Psi = 1$ for radiators and $\Psi = 0$ for floor heating). The heat delivered to the respective zones from the heat pump is
\begin{align}
  \Phi^h_t= f\n{cop}(40,T^a_t) \cdot P^e_t
\end{align}
where $P^e_t$ is the electrical power used by the heat pump. The solar irradiation $P^s_t$ is going through the windows and heating both the floor and the room. $\Psi_s$ describes the fraction that heats the room. 
%The Wiener processes, $d\omega\n{x}$, are used to describe a random diffusion of the temperature states, e.g. from occupant behavior and other activity in the building. How the parameter values are determined is presented later in Section \ref{Section:Buildings}.

\nomenclature[A]{$T$}{Temperature 
  \nomunit{$^{\circ}$C}}
\nomenclature[A]{$C$}{Heat capacity   
  \nomunit{$\text{kWh}/^{\circ}C$}}
\nomenclature[A]{$\Ps$}{Global solar radiation  
  \nomunit{$\text{kW}/\text{m}^2$}}
\nomenclature[A]{$R$}{Thermal resistance  
  \nomunit{$^{\circ}$C/kW}}
\nomenclature[A]{$\Ph$}{Heating 
  \nomunit{kW}}
\nomenclature[A]{$A$}{Area
  \nomunit{m$^2$}}
\nomenclature[A]{$g$}{Window glazing factor
  \nomunit{-}}
\nomenclature[A]{$\Psi_h$}{Binary variable defining the heating system (1 for radiators and 0 for floor heating)
  \nomunit{$-$}}
  \nomenclature[A]{$\Psi_s$}{Ratio of solar radiation in Interior vs. Floor.
  \nomunit{$-$}}
  
%\nomenclature[A]{$\omega$}{Wiener process describing the stochasticity of the model
%  \nomunit{$-$}}
%\nomenclature[A]{$\sigma$}{Scaling of the wiener process
%  \nomunit{$^{\circ}$C}}
\nomenclature[A]{$\text{COP}$}{Coefficient Of Performance
  \nomunit{$-$}}
\nomenclature[B]{$\n{f}$}{Floor zone
  \nomunit{}}
\nomenclature[B]{$\n{e}$}{Inner envelope zone  
  \nomunit{}}
\nomenclature[B]{$\n{a}$}{Ambient
  \nomunit{}}
\nomenclature[B]{$\n{i}$}{Interior zone 
  \nomunit{}}

\subsection{MPC}
The model is transformed from continuous into discrete time, see \cite{Madsen1995}. The discrete time linear state space model is written as
\begin{align}
        \boldsymbol{x}_{t} &=\boldsymbol{A}\boldsymbol{x}_{t-1} + \boldsymbol{B}u_{t-1} + \boldsymbol{E} \textbf{d}_{t-1} \\
        y &= \boldsymbol{C}\boldsymbol{x}_{t-1} + \epsilon_{t-1},
\end{align}
where $\boldsymbol{x}$ is the state vector (building temperatures) and $u$ is the controllable input vector describing is the electrical power to the heat pump, since the heat output from the heat pump, $\Ph$, is a function of both the power input signal and the COP factor. $\textbf{d}$ is the disturbances, which in this case is the outdoor temperature and solar irradiation. $y_t$ is thus the controllable variable $T^i_t$ plus some error $\epsilon_t$. Hence
\begin{equation}\label{eq:state}
    \boldsymbol{x}_t= \begin{bmatrix} T^i_t \\ T^f_t \\ T^w_t \end{bmatrix}  \; \; \; u_t = \begin{bmatrix}P^e_t\end{bmatrix} \; \; \; \boldsymbol{d}_t = \begin{bmatrix} T^a_t \\ G_t \end{bmatrix} 
\end{equation}

The matrix \textbf{A} states the dynamic behavior of the system, whereas matrix \textbf{B} specifies how the controllable input signals enter the system, and \textbf{E} specifies the uncontrollable input signals. Furthermore, $\boldsymbol{C}$ is a constant matrix that specifies the controllable state(s), in this case $\boldsymbol{C}=\begin{bmatrix} 1 & 0 & 0 \end{bmatrix}$. 
For a deeper explanation, please refer to: \cite{Halvgaard2012} and \cite{Madsen1995}.

The MPC then becomes a linear programming problem formulated as
\begin{equation} \label{eq:model}
\begin{array}{ll@{}ll}
\underset{u_s,v_k}{\text{arg min}}\;  & \displaystyle\sum\limits_{k=1}^{N} \lambda_{t+k}u_{t+k}+p_k v_{t+k}  &\\
\text{subject to}&\boldsymbol{X}_{s+1} \; = \;  \boldsymbol{A}\boldsymbol{X}_s + \boldsymbol{B}\boldsymbol{u}_s + \boldsymbol{E}\boldsymbol{D}_s \\
                 &T_{\text{min}} \; \leq \; \boldsymbol{C}\boldsymbol{X}_{s+1} + v_{s+1}\\
                 & T_{\text{max}} \; \geq \; \boldsymbol{C}\boldsymbol{X}_{s+1} - v_{s+1}\\
                 & 0 \; \leq \; u_s \; \leq  \; P_{\text{max}}\\
                 & v_s \; \geq \;  0\\
                 & \forall s \in \{t,t+1,t+2,...,t+N\} %\\                 & \boldsymbol{u}_s = \{u_{t+1},u_{t+2},...,u_{t+N}\}
\end{array}
\end{equation}
where $\lambda_{t+k}$ is the penalty at time $t+k$, which in this case is the marginal CO$_2$ emission intensity. $N$ is the prediction horizon. At each sampling time, the linear program is solved to obtain the heating schedule $[u_{t+1},u_{t+2},...,u_{t+N}]$. P\textsubscript{max} is the maximum power input signal the heat pump can receive.
As it may not always be possible to meet the temperature demand, a slack variable $v_k$ is introduced and connected to the violation penalty $p_v$. This value is set relatively high to avoid temperature violations. 

This linear program is solved using lpsolve interfaced with the R-package \textit{lpSolve}, \cite{lpsolve}.
\section{Inputs for the model}
\label{Section:Buildings}

In this section the reference buildings and input data are presented. 

The Danish building codes specify the building law requirements and contain the detailed requirements for all construction work.
This study evaluates the building codes from 1977 to 2018, denoted BC\textsubscript{year}. The requirements for outer walls, roof, windows and doors are listed by year in Appendix \ref{Appendix:BR}.

Two different types of buildings are considered; a family house and an office building, they differ in size and minimum temperature time settings during nightly setback. The night time set point is 18\textdegree{}C for [11pm:5am] and [6pm:7am] respectively for the family house and office building. During the day, the set point is 20\textdegree{}C.

For the sake of simplicity, the buildings are squares with one story. 
Typical building part constructions are used, illustrated in Appendix \ref{Appendix:parts}. The thickness of the concrete layers will be varied in the analysis. 

The construction material properties are listed in Table \ref{tab:materials} in Appendix \ref{Appendix:Materials}, where also the building dimensions and model parameters are defined. 

The windows are defined as equal sizes on each side of the house pointing in north, east, south and west respectively with a window-to-wall ratio of 0.11 (the proportion of the wall that is windows). The \textsf{R}-package \textit{solaR} is used for solar radiation inclination angle calculations.
$\Psi_s$ is set to 0.1 as in \cite{Halvgaard2012}.

Dimensioning of the heat pump is based on the heat loss from the buildings, specified in Appendix \ref{Appendix:Materials}.

\subsection{Forecasts}\label{section:Forecast}
24 hour horizon CO$_2$ emission forecasts presented in the related paper, \cite{KL}, are used. The real time values are presented in Sec. \ref{sec:data} along with estimated real time values of the temperature and solar radiation. The weather forecasts have horizons of 24 hours too.

To evaluate the MPC and the impact of using forecasts, different extreme cases are defined:

\begin{itemize}
    \item \textbf{Case\textsubscript{Ideal}}: This takes the exact value of a future CO$_2$ emission intensity as prediction hence, a perfect forecast. This provides an upper limit of CO$_2$ savings.
    \item \textbf{Case\textsubscript{Real}}: This takes the CO$_2$ emission forecast developed in \cite{KL} and represents the performance of the MPC with real forecasts.
    \item \textbf{Case\textsubscript{Trivial}}: This makes no use of forecasts and will thus result in a non predictive controller that simply controls the heat pump keeping the temperature at the lower limit if possible.
\end{itemize}

\iffalse
%Real time weather measurements for the evaluated period are not included as this could lead to non-representative results unless a sufficient amount of data points are considered, and these have not been accessible to the author. If local measurements are accessible, the weather forecasts can be corrected to the local measurements. 
Observed weather data is not used in the analysis. Using observed data from a particular location would introduce a particular effect of the quality of the weather forecasts for this location. Therefore, the 1 to 6 hour horizon forecast is assumed to be the real time data, with only the modification of smoothing gaps in the solar radiation, as described in ??. It implies that every horizon shorter or equal to 6 hours, are actually assumed to be nearly perfect weather forecasts. The COP for the heat pump is varying with the temperature and is also included in the model as a COP forecast.
\fi
\begin{figure*}[t]
\centering
  \includegraphics[width=\linewidth]{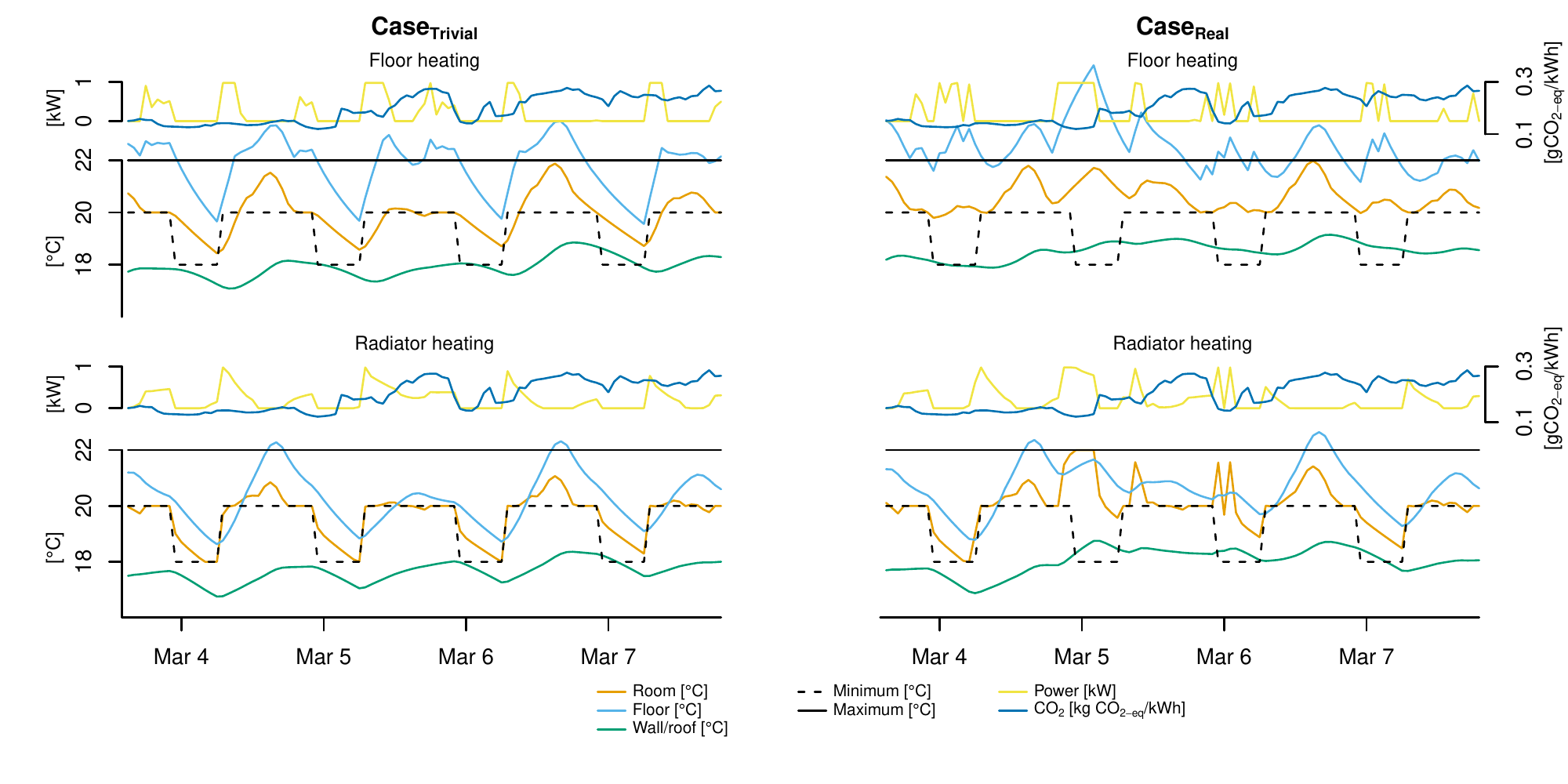}
  \caption{Varying temperature constraints - night [11pm:5am]; 18\textdegree{}C. Note the CO2 emission intensity and heating does not follow the Y-axis range, but rather the specified range in the colour legend. }
  \label{fig:fam_floor}
\end{figure*}

\section{Results} \label{section:Results}
\small

In this section, various conditions and parameters are evaluated, e.g. effect of horizon length, heat pump size, insulation and concrete thicknesses. The radiator and floor heating system is compared throughout the analysis along with the family house versus the office building.
The criteria to be optimized is the CO$_2$ emission savings. The total emissions are calculated from:
$ \Lambda(\text{Case}) = \sum_{t=1}^{n-N} u_{\text{Case},t} \lambda_t $,
where $n$ is the number of data points presented in Sec. \ref{sec:data} ($n=8688$)  and the 'Case' denotes one of the three cases defined in Sec. \ref{section:Forecast}. The savings are thus calculated by
\begin{equation} \label{eq:savings}
    \mathit{Savings(Case)} = \frac{ \Lambda(\text{Case\textsubscript{Trivial}}) - \Lambda(\text{Case})}{ \Lambda(\text{Case\textsubscript{Trivial}})}
\end{equation}
As previously noted, this measure indicates the relative savings from utilizing the energy flexibility, hence not the absolute savings. If not otherwise stated the results are calculated with a building complying with BC\textsubscript{2018}, see \ref{Appendix:BR}.

\iffalse
The parameters that will be investigated are:

\begin{itemize}
    \item \textbf{Heating system and varying set points:} Both radiator and floor heating are considered and the use of varying set points (lower temperature during the night). 
    \item \textbf{Horizon of forecasts:} To get an idea about how long horizons actually are needed to get a well performing MPC. 
    \item \textbf{Size of heat pumps:} Essential for comparing the buildings and to know whether the potential is reached. Also economically, this is important because as the price increases with larger heat pumps. This will become a compromise between price and CO$_2$ emission. The default values for the family house and office are the minimum sizes required to meet the heat demand on the coldest day (-12 \textdegree{}C); 3 and 13 kW\textsubscript{heat} respectively (see \ref{Appendix:Materials} for calculations). Requirements are thus a 1 and 4.3 kW input signal respectively according to Equation \ref{eq:COP}. 
    \item \textbf{Insulation and concrete thicknesses:} These will be adjusted to see the impact of levels of insulation and heat capacity. The default thicknesses and material properties are shown in \ref{Appendix:Materials}.

\end{itemize}

\fi
\renewcommand{\arraystretch}{1}

An example of the differences between the cases, and the radiator and floor heating, is illustrated in Figure~\ref{fig:fam_floor}. The resulting electrical power and temperature for Case\textsubscript{Trivial} and Case\textsubscript{Real} on a four day period for the family house is shown. The result of Case\textsubscript{Trivial} is slightly different for the two heating systems. With radiators, it needs to heat more continuously than the floor heating throughout the day. This is because the radiators transfers the heat directly to the internal air, and not through the large heat capacity in the floor, resulting in a much faster response. In both cases the heating is switched off during the night time to reach the lower set point. However, the floor heating violates the temperature restrictions more during the morning while heating up the house, which is due to its slow response. Case\textsubscript{Real} seeks to only switch on the heat pump during low emission periods. The radiator system does this well, but it is clearly limited by the maximum indoor temperature limit and the power input decays immediately to avoid temperature violation. In the floor heating system, the heat pump can operate at full load for longer time using the floor as storage. An interesting point is that using day and night profiles, Case\textsubscript{Real} has no benefits of letting the temperature drop during the night because of: i) the temperature response is too slow and ii) the emissions are usually lowest during the night, so this is the best time to use the heat pump. Contrarily, the indoor temperature in the radiator system occasionally drops during the night if there is no significant drop in CO$_2$ emissions. 

As expected, over the course of the period of almost a year ($n = 8688$ hours) the floor heating system provides slightly more flexibility and reaches savings of 11\% against 9\% using radiators for the whole year for Case\textsubscript{Real}.

\subsection{Control Horizon}

\begin{figure}[!tbp]
  \centering
  \begin{minipage}[b]{0.49\textwidth}
  \includegraphics[width=\linewidth]{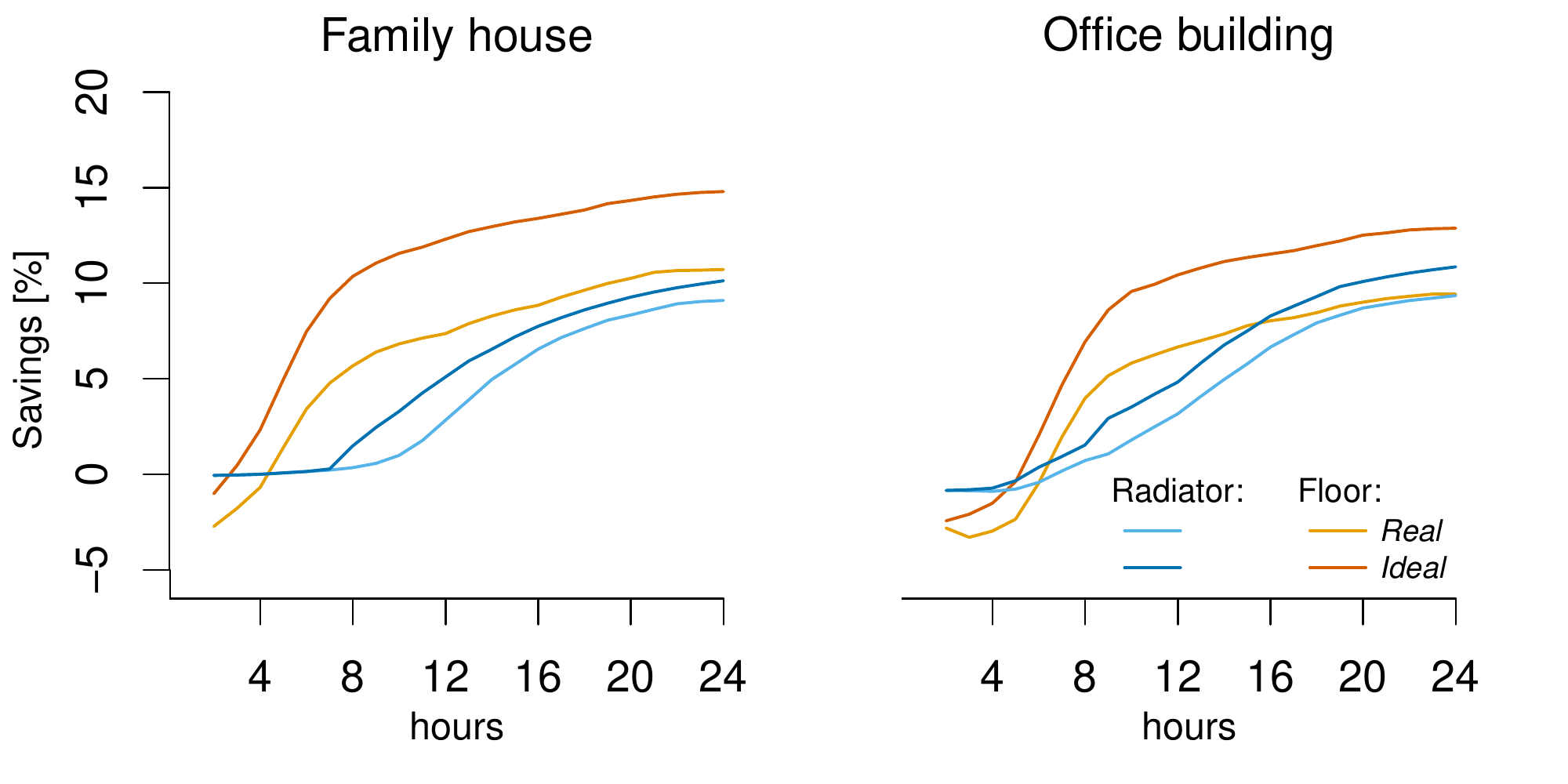}
  \caption{Savings with respect to the forecast horizon. Shown for both an ideal (perfect) and the Real forecasts.}
  \label{fig:horizon}
  \end{minipage}
  \hfill
  \begin{minipage}[b]{0.49\textwidth}
  \includegraphics[width=\linewidth]{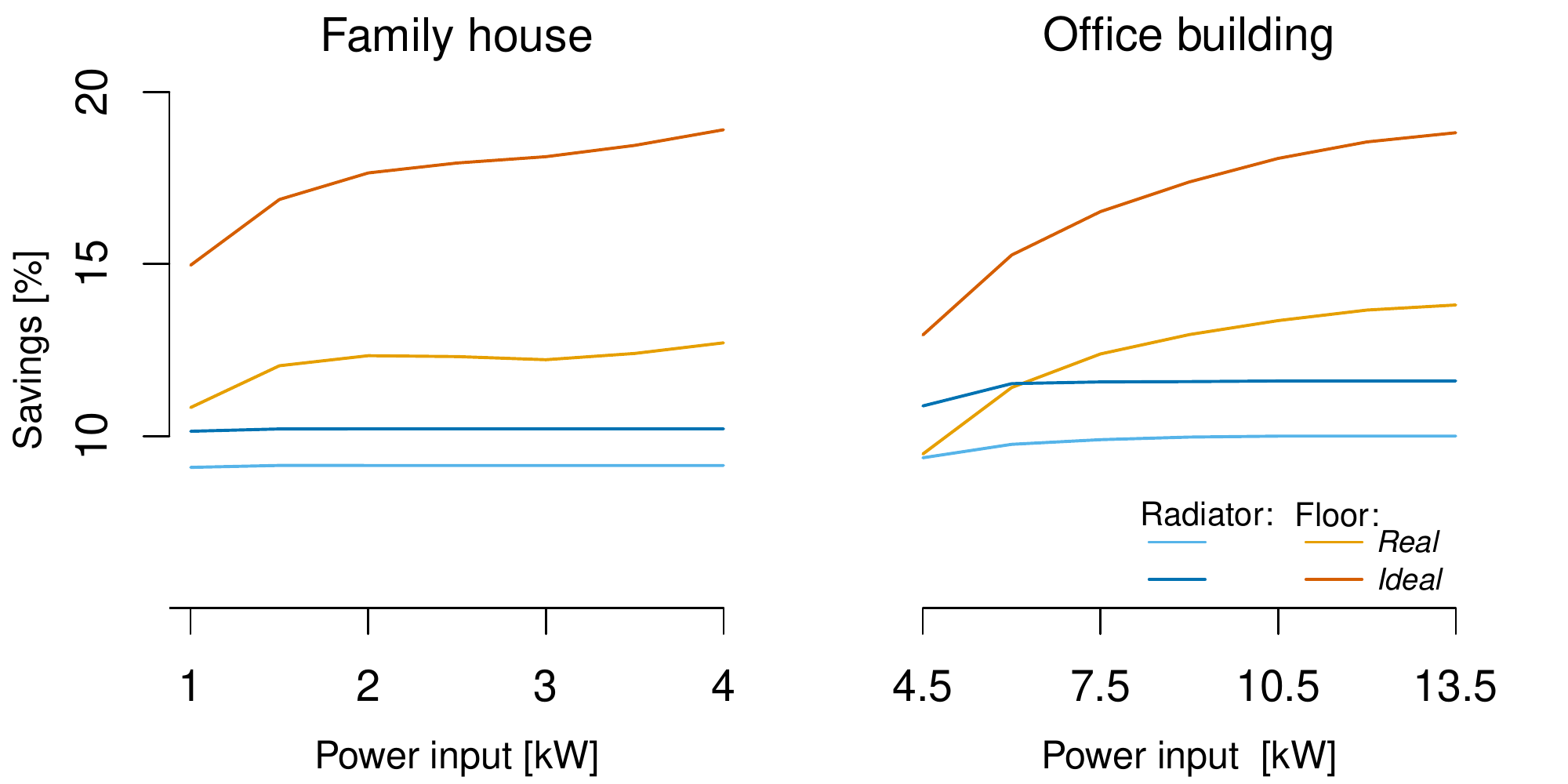}
  \caption{CO\textsubscript{2} emission reduction as a function of the size of the heat pump in kW shown for the real forecasts.}
  \label{fig:maxp}
  \end{minipage}
\end{figure}

The control horizon needs to be sufficiently long for the MPC to provide flexibility to the system. In Figure~\ref{fig:horizon}, the savings are plotted versus the control horizon in the different scenarios. Note, the savings from floor heating become negative when using low control horizons. That is caused by the nightly setback; Case$_{Trivial}$ switches on at six AM every morning and the emission peak is happening already at four AM, see Figure~\ref{fig:shift} showing the average switch in hourly demand alongside the average emissions. When using e.g. a two hour control horizon, the heat pump will be forced to switch on at four AM instead and thus increase the emissions.

 Interesting to note is the changing behaviour of the curves around the 8 hours horizon: Case$_{Ideal}$ in the family house with radiators has no savings up until this point. Like any other energy storage a loss is introduced, in this case by an increase in temperature resulting in a higher heat loss. Therefore, the MPC will only store heat if the CO$_2$ variations are large enough for the resulting emissions to break even with the increased losses. This is less of a problem for the floor heating system, as loss is much lower. This behaviour is less pronounced in the office building with radiators, because the building is larger, hence it has a higher heat capacity.

 The loss in savings due to forecast errors, can be found from the difference between \textit{Ideal} and \textit{Real}. The radiator system is close to reaching its full potential, where the floor heating still can improve maximum about 5\% savings from better forecasts. 

Finally, it is noted, that there is still an increase in savings at the 24 hours horizon, for all scenarios, indicating that even longer horizons will lead to further increase in savings.

\subsection{Size of heat pump}
The savings as functions of the heat pump size are shown in Figure \ref{fig:maxp}. For the radiator systems, there is no significant gain in increasing the heat pump size for neither the family house or the office building. For the family house with floor heating, there is a relatively large gain when considering Case$_{Ideal}$, while Case$_{Real}$ only reaches a slight improvement. There are significantly higher savings to reach from Case$_{Real}$ in the office building; 
because of the larger floor to wall area ratio (more heat capacity relative to the area the heat can escape through), a larger heat pump  can accumulate more heat and thereby increase the flexibility.

\subsection{Insulation and concrete thickness}

\begin{figure*}[t]
\centering
  \includegraphics[width=\linewidth]{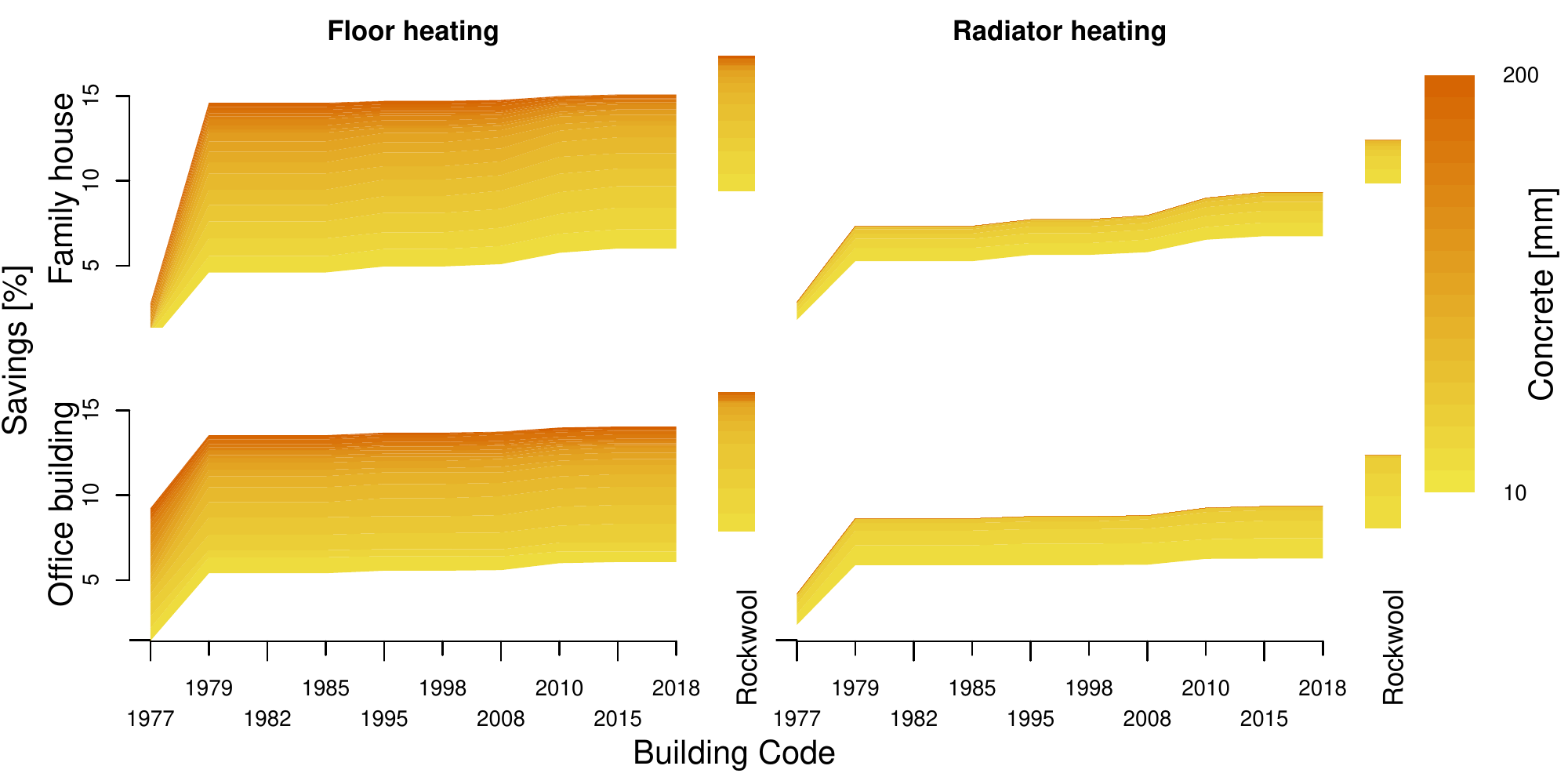}
  \caption{Savings with respect to the building code year and concrete thickness in the floor. Note, "Rockwool" represents the, by Rockwool A/S, recommended U-values for walls and roof (doors and windows correspond to BC\textsubscript{2018}). Refer to Appendix \ref{Appendix:BR}.).}
  \label{fig:BR}
\end{figure*}

The savings are evaluated with respect to the development of building codes from 1977 and forward, which has been an increase in insulation, window and door requirements (see Appendix \ref{Appendix:BR} for building code specs and corresponding physical values). Another important aspect is the concrete thickness in the floor because it increases the heat capacity and thus the heat storage capabilities. In Figure~\ref{fig:BR} the savings are shown as a function of the building codes and concrete thickness.

Generally, BC$_{1977}$ houses have very little potential, however, in large buildings (office buildings) this increases to around 9\% savings when 200 mm of concrete is added to the floor. Also, the savings in the office building are less sensible to the building codes. Both conclusions have to do with the floor to wall ratio. The larger it is, the more the concrete thickness in the floor can contribute, and the less the insulation in the walls contributes.

As seen in the figure for radiator heating, the concrete thickness in the floor is less important for the savings. Still, adding 40 mm can increase the savings by around 3\%, but any thicker layer will not increase the savings at all (i.e.\ the red colored area in the graphs for radiator heating can hardly be seen). The insulation thickness in the office building using radiators seems saturated after BC$_{1979}$. This may seem counter intuitive, but in reality with high thermal resistance, the model becomes more rigid and therefore less flexibility is allowed to occur. Of course, the absolute CO\textsubscript{2} emission savings will increase with more insulation.

Floor heating provides most flexibility in both buildings given the concrete in the floor is thick (around 14\%-15\% in both with 200 mm concrete and BC$_{2018}$). With only 10 mm of concrete, the savings are almost identical for radiator and floor heating (about 6\% with BC$_{2018}$ in all cases). Note that a concrete thickness of 10 mm is not common practise.

Using the Rockwool recommendations, the savings increase to around 17\% with floor heating for both buildings (200 mm concrete).

\section{Discussion}
\label{Section:Discussion}
\small

Buildings built according to BC$_{1979}$ or later can expect savings of 4\%-17\% depending on various factors e.g. heating system, level of insulation and floor concrete thickness. However, if built according to BC$_{1977}$, the savings are only expected to be 0-9\%. This study only measured the relative savings, but the absolute savings will of course also increase with more insulation. Therefore, if a house is not well insulated, the first attempt to lower CO$_2$ emissions should be to lower the heat demand by adding insulation before trying to optimize the control.
As a result of new regulations in BC$_{1979}$, buildings in Denmark built prior to 1979 have been increasingly re-insulated to decrease the heat demand and costs. In \cite{Broegger}, it is found that the actual heat demand is on average lower than the theoretical calculations based on registered building data. This could imply that buildings indeed are re-insulated without further registration.

Most buildings in Denmark are built prior to 1979, \cite{dst}, and this group is therefore the best representative for the potential buildings - floor heating is rarely the main heating system in this group. It would thus follow the early end of the radiator curves in Figure~\ref{fig:BR} with re-insulated buildings likely to comply with BC$_{1979}$. Therefore between 4\% and 7\% of savings can be expected. However, new buildings following Rockwool recommendations and have floor heating installations reach savings of nearly 17\% using MPC. Often, buildings combine radiators and floor heating and rely not only on one or the other. Thus in reality the advantages from both systems can be utilized. Recall from Figure~\ref{fig:fam_floor}, the radiators are good at quick responses and therefore allows the temperature to decrease during the night contrarily to the floor heating. However, it has little capability of storing heat for longer periods. The potential for this is open to further studies.

\begin{figure}[t]
\centering
  \includegraphics[width=\linewidth/2]{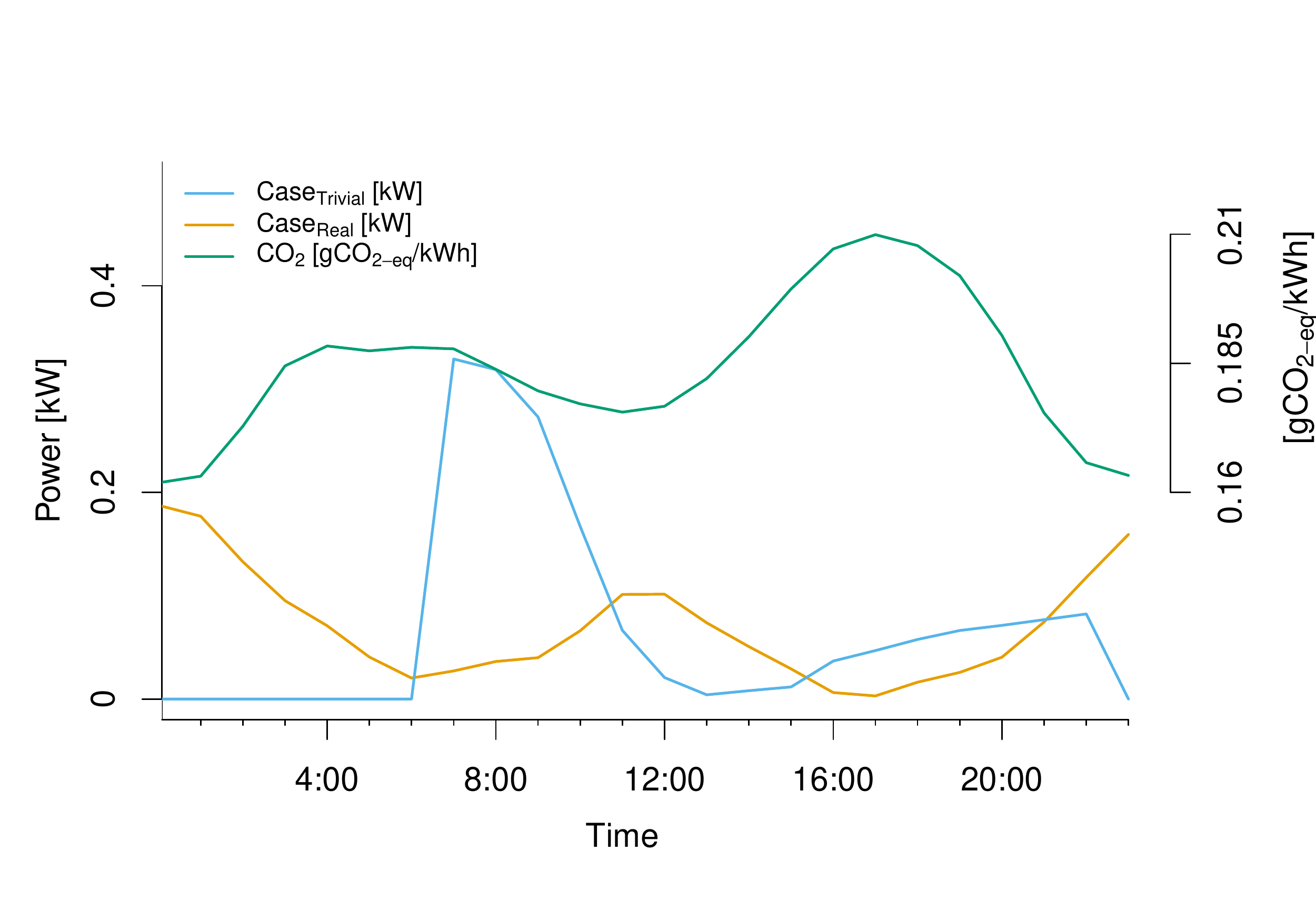}
  \caption{Heat pump load on average as function of the hour of the day for both Model$_{Trivial}$ and Model$_{Real}$. Hourly average marginal CO$_2$ emissions are shown in green.}
  \label{fig:shift}
\end{figure}

In Figure~\ref{fig:shift}, the average daily load shift for a BC$_{2018}$ family house is shown together with the marginal CO$_2$ emissions. 
The trivial control follows to a certain degree the emissions throughout a day. The natural need for heating is therefore very inconvenient for the emissions and illustrates why energy flexibility is important. The MPC smooths the load during the day, decreasing it at otherwise peak hours and shifts most of the load to operate at midnight despite the lower temperature set point. 
This illustrates the importance of predictive control - if all houses follows the same schedule, there will be a need for much more additional storage capacity.

 This study is specifically for conditions in DK2 for a one year period in 2017 and 2018 - 36\% wind and solar power production share. The results will change depending on the conditions e.g. with higher levels of renewable fluctuating generation in the future \cite{ea}. In a 100\% renewable scenario it is of course not meaningful to use the CO$_2$ emissions as minimizing objective, however utilizing energy flexibility will be vital\cite{LUND2009524}.

\subsection{Model simplification}
The results are based on calculations using a simplified building model of a simplified building with only a single room. If more rooms are considered, it would add more heat capacity from walls dividing the rooms. This could increase the savings. Furthermore, a storage tank could have been added, which also could increase the savings as it would boost the energy flexibility.

Disturbances, other than solar radiation and ambient temperatures, have been neglected in this study, e.g. human activity. It is left as a point for future studies to include and assess the impact of building usage in the models and analysis. 
%It can be estimated through real temperature measurements and in a real implementation, it can use adaptive machine learning techniques to learn the human activity behavior for a given building. Experiments have been performed using some values of white noise standard deviations and no significant changes in the savings are seen with standard deviations lower than 0.3 \textdegree{}C.

%The noise in the model is uniform along the time series, where in reality, it would be a fair assumption that this noise varies during the day. During the night, it may be low due to low human activity, whereas it in the evening might be high due to dinner preparation and other family activities. 

% The study is limited to regional weather data, so in a real implementation local weather observations should be acquired, and local forecasts can be created by correcting the regional weather forecast to the local one. 

% Another important aspect is the fact that the heat pump efficiency is not only depending on the outside temperature, but also on the relative load. This model is simplified to only consider the outside temperature as a factor for the efficiency.

% 
\section{Conclusion}\label{Section:Conclusion}
\small
The potential of achieving CO$_2$ savings using the energy flexibility of buildings has been analysed in a range of relevant scenarios including both a family house and an office building. The simulated buildings were heated with heat pumps and the characteristics were varied according to the historic development of Danish building codes. The CO$_2$ level of the electricity generation for a year long period in the Danish area DK2 was used as input, together with weather data from the area. 

The results show considerable CO$_2$ savings for both radiator and floor heating systems. Forecast horizons should be 24 hours or longer to obtain the full savings potential. Over-dimensioning the heat pump to increase flexibility turns out only to yield significant savings with floor heating, especially in the office building due to its larger storage area relative to the wall area.

Following the Rockwool insulation recommendations, savings can reach around 17\% using floor heating with 100 mm of floor concrete and 14\% using radiators. Generally, buildings must comply with building codes later than 1977 to achieve any considerable savings due to the low requirements in BC$_{1977}$. 

Predictive control is vital to eliminate the peak hours, especially happening in the morning after the nightly setback, which is common practise. It is able to shift the demand to periods, where the coal power production is low, typically out of cooking peaks and during night time.

%\subsection{Future work}
%This study uses a simple model, where users are not taken into account.
%If time and resources were limitless, an MPC should be implemented in one or more test houses with representative families living a normal everyday life to acquire real world data, with sensors monitoring climate and penalty (CO\textsubscript{2}). Then the problems with no users, non representative houses/buildings had been eliminated.

% Only space heating is considered. However, implementing MPC to hot water heating is also valid. Especially with the introduction of learning that can learn consumer habits and improve the heat demand forecast. Further studies need to be conducted to get further insight in this and to what extend it may require larger water tanks for storage.
 
% Also reversible heat pumps could be interesting to focus on because this allows for operation during the summer too. 

%%%%%%%%%%%%%%%%%%%%%%%%%%%%%%%%%%%%%%%%%%
\vspace{6pt} 

%%%%%%%%%%%%%%%%%%%%%%%%%%%%%%%%%%%%%%%%%%
%% optional
%\supplementary{The following are available online at \linksupplementary{s1}, Figure S1: title, Table S1: title, Video S1: title.}


%%%%%%%%%%%%%%%%%%%%%%%%%%%%%%%%%%%%%%%%%%
\authorcontributions{conceptualization, R. Grønborg and P. Bacher; methodology, K. Leerbeck and A. Tveit; software, R. Grønborg and K. Leerbeck; validation, K. Leerbeck; formal analysis, K. Leerbeck; investigation, K. Leerbeck; resources, R. Grønborg, P. Bacher and H. Madsen; data curation, K. Leerbeck. and O. Corradi; writing--original draft preparation, K. Leerbeck; writing--review and editing, R. Grønborg, P. Bacher, O. Corradi and H. Madsen; visualization, K. Leerbeck; supervision, R. Grønborg and P. Bacher; project administration, P. Bacher, H. Madsen; funding acquisition, P. Bacher, H. Madsen}

%%%%%%%%%%%%%%%%%%%%%%%%%%%%%%%%%%%%%%%%%%
\funding{This research was supported through the project “Smart Cities Accelerator 2016-2020”  funded by the EU program Interreg Öresund-Kattegat-Skagerrak, the European Regional Development Fond and the CITIES project (DSF1305-00027B).
}

%%%%%%%%%%%%%%%%%%%%%%%%%%%%%%%%%%%%%%%%%%
\acknowledgments{We are thankful for Tomorrow \footnote{\url{www.tmrow.com}} who has provided the data used in this study (including emission calculations for the bidding zone DK2). }

%%%%%%%%%%%%%%%%%%%%%%%%%%%%%%%%%%%%%%%%%%
\conflictsofinterest{The authors declare no conflict of interest. The funders had no role in the design of the study; in the collection, analyses, or interpretation of data; in the writing of the manuscript, or in the decision to publish the results.} 

%%%%%%%%%%%%%%%%%%%%%%%%%%%%%%%%%%%%%%%%%%
%% optional
\abbreviations{The following abbreviations are used in this manuscript:\\

\noindent 
\begin{tabular}{@{}ll}
MDPI & Multidisciplinary Digital Publishing Institute\\
CHP & Combined Heat and Power \\
MPC & Model Predictive Control 
\end{tabular}}

%%%%%%%%%%%%%%%%%%%%%%%%%%%%%%%%%%%%%%%%%%
%% optional
\appendixtitles{no} %Leave argument "no" if all appendix headings stay EMPTY (then no dot is printed after "Appendix A"). If the appendix sections contain a heading then change the argument to "yes.
\appendix
\section{Kernel smoothing}\label{Appendix:kernel}
The 1 to 6 hours horizon solar radiation forecasts are smoothed in order to remove large shifts in value, thus providing a better representation of real conditions. 

 Assuming horizons for $h=1$ are the most accurate forecasts, a kernel smoothing process using a weight for short horizon favouritism is applied. The kernel weight, $\mathbf{w}_{1}$, is defined as the Epanechnikov kernel;
 \begin{align}
     \mathbf{w}_{1} = \frac{3}{4} (1-\mathbf{u}^2),
 \end{align}
 where $\mathbf{u}= \frac{|\text{x}_i - \mathbf{x}|}{b}$. $\mathbf{x}$ is a vector $[1,2,..,n]$, where $n$ is the number of data points and $b$ is the bandwidth. The short horizon favoritism weight is defined as 
 \begin{align}
     \mathbf{w}_{2} = \mathrm{e}^{-\frac{a}{\mathbf{h}-1}},
 \end{align}
 where $\mathbf{h} \in [1,2,...,6]$ represents the hours in advance to the observation the forecast was received. Together, $\mathbf{w}_1$ and $\mathbf{w}_2$ define the final weight function; $\mathbf{w} = \mathbf{w}_1\cdot \mathbf{w}_2$. This is applied into a linear regression model
\begin{align}\label{eq:linear}
    \mathbf{y} &= \mathbf{X}\boldsymbol{\beta} + \epsilon, \\
    \text{for} \; \epsilon &\sim N(0,\sigma^2 I), \nonumber
\end{align}
where $\boldsymbol{X}$ is the input matrix (explanatory variables; hour, day and month), $\boldsymbol{y}$ is the output vector (response variable: Solar irradiation) and $\boldsymbol{\beta}$ is a vector of regression coefficients to be found. $\epsilon$ represents the normally distributed errors in the model. 

\begin{figure}[!tbp]
  \centering
  \begin{minipage}[b]{0.49\textwidth}
  \includegraphics[width=\linewidth]{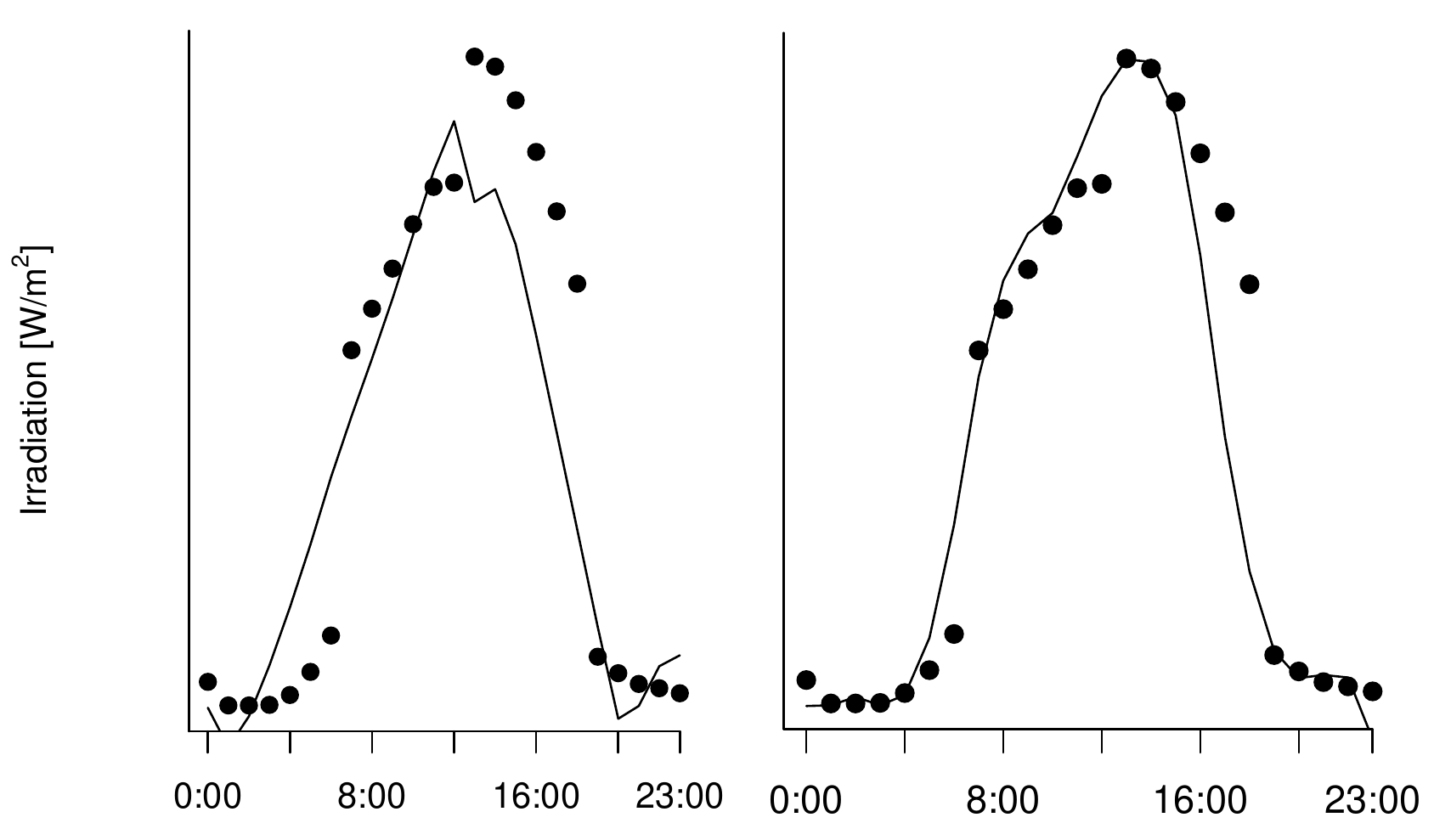}
  \caption{i) Solar irradiation 6 hour horizon forecasts for the 21st of June 2017 (updated at 2am,8am,14pm and 20pm) and estimates of the real time irradiation (solid line) w/o splines (left). ii) Estimates of the real-time irradiation using splines and the kernel smoothing approach (right)}
  \label{fig:kernel}
  \end{minipage}
  \hfill
  \begin{minipage}[b]{0.49\textwidth}
  \includegraphics[width=\linewidth]{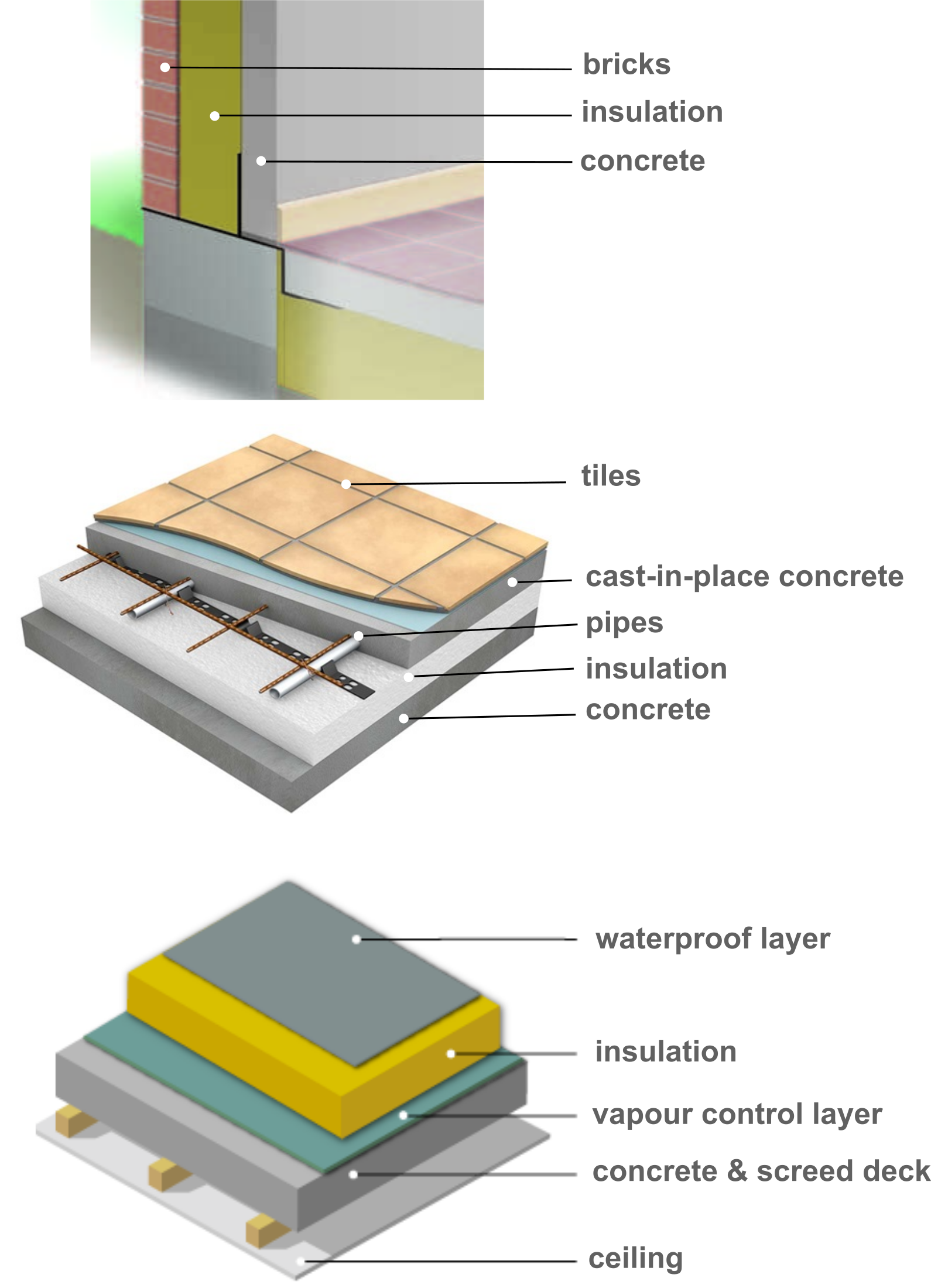}
  \caption{Typical constructions of walls  (top), floor  (middle) \cite{Upo1} and roof (bottom) \cite{Greenspec}.}
  \label{fig:roof}
  \end{minipage}
\end{figure}

The least square regression is performed to minimize
$$ S(\boldsymbol{\beta}) =  ||\mathbf{y} - \mathbf{X}\boldsymbol{\beta}||^2 $$
and obtain the weighted least-squares solution
\begin{equation}
\boldsymbol{\beta} = (\mathbf{X}^T \mathbf{w} \mathbf{X})^{-1} \mathbf{X}^T \mathbf{w} \mathbf{y}.\label{eq:coeff}
\end{equation}

where $\mathbf{W}$ is the weight vector $w$. Using $a = 1.5$ and $b = 7$, the results are illustrated in Figure \ref{fig:kernel} (left plot). This is not sufficient on its own, as it does not manage to capture the curvature and midday peak. Therefore, the hour is converted into base splines (local polynomials between specified points called knots \cite{Springer}); 
\begin{equation} 
 \mathit{bs}(\mathbf{hour}_t) =
 \begin{bmatrix} \mathit{bs}_0(\mathbf{hour}_t) & \mathit{bs}_1(\mathbf{hour}_t) &  \ldots  & \mathit{bs}_{df}(\mathbf{hour}_t)
 \end{bmatrix}^{\top}.
 \end{equation}
 where $\mathbf{hour}_t$ is the hour corresponding to the time step $t$. $df$ is the degrees of freedom (essentially the number of splines; the higher the better it will fit the actual values). Using $df = 7$, the final result is illustrated in Figure \ref{fig:kernel} (right plot).

%In order to estimate the $\beta$ coefficients, the errors are minimized, so the objective function to be minimized becomes;
%$$ S(\beta) =  Y - \beta X $$
%And to put it as least square regression, it is; $ S(\beta) =  ||y - \beta X||^2 $ \\
%The solution for this problem is given as the ordinary least-squares estimator;
%$$ \beta_{ols} = (X^T X)^{-1} X^T y $$
%The proof for this will not be discussed here. 

\section{Building regulation data}\label{Appendix:BR}

\begin{table}[htbp]
  \centering
  %% Variable for scaling the tikz figure
  \newcommand{\scaleTikz}{0.45}
  %% Control text size
  % \large
  %\usetikzlibrary{decorations.pathreplacing}

\begin{tikzpicture}[t]

\node  at (-150pt,-100pt) {\resizebox{1\hsize/2}{!}{

\begin{tabular}{ l | l  l  r |  r  r  r }

	 \textbf{BR} & \textbf{Walls} & \textbf{Roof} & \textbf{Doors} & \multicolumn{3}{c}{\textbf{Windows}}  \\ 
	 \textit{Year} &  $U \left[ \frac{\text{W}}{\text{m\textsuperscript{2}} \text{K}} \right]$ &  $U \left[ \frac{\text{W}}{\text{m\textsuperscript{2}} \text{K}} \right]$ &  $U \left[ \frac{\text{W}}{\text{m\textsuperscript{2}} \text{K}} \right]$ &  $U \left[ \frac{\text{W}}{\text{m\textsuperscript{2}} \text{K}} \right]$ & $E_{ref} \left[ \frac{\text{kWh}}{\text{m\textsuperscript{2}}} \right]$ & $g [-]$  \\ [5pt] \hline
	 
	 1977 & 1 & 0.45 & 3.6 & 3.6 & -174-314* & 0.777* \\ 
	 1979 & 0.4 & 0.2 & 2 & 2.9 & -117.378* & 0.744* \\
	 1982 & 0.4 & 0.2 & 2 & 2.9 & -117.378* & 0.744* \\
	 1985 & 0.4 & 0.2 & 2 & 2.9 & -117.378* & 0.744* \\
	 1995 & 0.4 & 0.2 & 2 & 2.3 & -69.934* & 0.709* \\
	 1998 & 0.4 & 0.2 & 2 & 2.3 & -69.934* & 0.709* \\
	 2008 & 0.4 & 0.2 & 2 & 2 & -46.909* & 0.688* \\
	 2010 & 0.3 & 0.2 & 2 & 1.8* & -33 & 0.673* \\
	 2015 & 0.3 & 0.2 & 2 & 1.6* & -17 & 0.654* \\
	 2018 & 0.3 & 0.2 & 2 & 1.6* & -17 & 0.654* \\
	 Rockwool A/S & 0.14 & 0.1 & - & - & - & -
\end{tabular}

}

};
\end{tikzpicture}
  \caption{Building regulation data by year. *Estimated values: From 2010, the windows insulation properties are described by $E\n{ref}$, estimated net heat transfer through the window into the room (follows; $E\n{ref} = 194.4 g - 90.36  U$). An exponential relationship between $U$ and the glazing ($g$) is found from key numbers of different window types to be; $U=0.0205 e^{6.6545 g} $, \cite{more_windows}. This essentially allows $U$ and $g$ to be calculated from $E\n{ref}$.}
  \label{fig:BR_tab}
\end{table}

\FloatBarrier
\section{Building parts}\label{Appendix:parts}
The buildings consist of three parts; the walls, the floor and the ceiling. The material composition in each part is illustrated in Figure \ref{fig:roof}. Thereto, the model parameters (heat capacities and thermal resistances) are defined in Figure \ref{fig:parameters}. The wall and roof are divided into the inner and outer part where only the inner temperature is modelled. The inner part is everything between the room and the insulation.

\section{Parameter estimation and heat pump dimensions}\label{Appendix:Materials}

The model parameters are defined illustratively in Figure \ref{fig:parameters}.
The materials used and corresponding properties are listed in Table \ref{tab:materials}.

\begin{figure}[!tbp]
  \centering
  \begin{minipage}[b]{0.49\textwidth}
  \newcommand{\scaleTikz}{0.45}
  %% Control text size
  \large
  \input{Graphics/parameters.tex}
  \caption{Definition of model parameters (heat capacities and thermal resistances) in the roof (left), floor (middle) and wall (right).}
  \label{fig:parameters}
  \end{minipage}
  \hfill
  \begin{minipage}[b]{0.49\textwidth}
  %% Variable for scaling the tikz figure
  \newcommand{\scaleTikz}{0.45}
  %% Control text size
  \large
  \usetikzlibrary{decorations.pathreplacing}

\begin{tikzpicture}[t]

\node [rotate=90,anchor=north] at (-390pt,-63pt) {\resizebox{.05\hsize}{!}{Wall}};
\draw [decorate,decoration={brace,amplitude=5pt,raise=4pt},yshift=-83pt]
(-370pt,0) -- (-370pt,1.3);

\node [rotate=90,anchor=north] at (-390pt,-40-67pt) {\resizebox{.05\hsize}{!}{Roof}};
\draw [decorate,decoration={brace,amplitude=5pt,raise=4pt},yshift=-83pt-48pt]
(-370pt,0) -- (-370pt,1.48);

\node [rotate=90,anchor=north] at (-390pt,-40-67-45pt) {\resizebox{.05\hsize}{!}{Floor}};
\draw [decorate,decoration={brace,amplitude=5pt,raise=4pt},yshift=-83pt-87pt]
(-370pt,0) -- (-370pt,1.1);

\node  at (-260pt,-100pt) {\resizebox{1\hsize}{!}{

\begin{tabular}{ |>{\small} l |>{\small} l | >{\small} l | >{\small}r |>{\small} r | >{\small} r | >{\small} r | >{\small} r | }
\hline
	& Index & \textbf{Material} & $\boldsymbol{\zeta}$ [m] & $\boldsymbol{\rho} [\frac{\text{kg}}{\text{m}^3}] $ & $\boldsymbol{C} [\frac{\text{J}}{\text{kg}\cdot \text{K}}]$ & $\boldsymbol{k} [\frac{\text{W}}{\text{m}\cdot \text{K}}]$ & $\boldsymbol{R} [\frac{\text{m}^2\cdot \text{K}}{\text{W}}]$ \cr  \\ \hline
	\textbf{Surface, outer} & w,s,o & - & -  & -  & -  & -  & 0.06  \\ \hline
	\textbf{Outer} & w,o & \textbf{Bricks} & 0.15 & 1920 & 790 & 0.9 & 0.167 \\ \hline
	\textbf{Insulation} & w,insul & \textbf{Rockwool} & 0.12 & 240 & 710 & 0.042 & 2.693 \\ \hline
	\textbf{Inner} & w,i &  \shortstack[l]{\textbf{Concrete}, \\ \textbf{light weight}} & 0.10 & 1600 & 840 & 0.79 & 0.127 \\ \hline
	\textbf{Surface, inner} & w,s,i & - & -  & -  & -  & -  & 0.12  \\ \hline
	\  & \ & \  & \  & \  & \  & \  &  \\ \hline
	\textbf{Surface, outer} & r,s,o & \textbf{Waterproof layer} & 0.01 & 0 & 0 & 0 & 0.06 \\ \hline
	\textbf{Insulation} & r,insul & \textbf{Rockwool} & 0.25 & 144 & 1000 & 0.058 & 4.304 \\ \hline
	\textbf{Air space} & r,air & \textbf{Air} & 0.05 & 1.225 & 1000 & 0.026 & 0.400 \\ \hline
	\textbf{Concrete} & r,con & \shortstack[l]{\textbf{Concrete}, \\ \textbf{light weight}} & 0.05 & 1600 & 840 & 0.79 & 0.063\\ \hline
	\textbf{Ceiling} & r,cei & \textbf{Plaster light} & 0.01 & 1680 & 840 & 0.81 & 0.012 \\ \hline
	\textbf{Inner surface} & r,s,i & - & -  & -  & -  & -  & 0.16  \\ \hline
	\ & \ & \ &  \  & \  & \  & \  &   \\ \hline
	\textbf{Inner surface} & f,s,i &- & -  & -  & -  & -  & 0.11  \\ \hline
	\textbf{Cover} & f,cov & \textbf{Plywood} & 0.01 & 545 & 1210 & 0.12 & 0.083 \\ \hline
	\textbf{Concrete} & f,con & \shortstack[l]{\textbf{Concrete}, \\ \textbf{light medium}} & 0.05 & 1600 & 840 & 0.79 & 0.063 \\ \hline
	\textbf{Insulation} & f,insul & \textbf{Rockwool} & 0.30 & 240 & 710 & 0.042 & 7.143 \\ \hline
\end{tabular}
}
};

\end{tikzpicture}
  \caption{Thickness ($\boldsymbol{\zeta}$), density ($\boldsymbol{\rho}$), heat capacity ($\boldsymbol{C}$), thermal conductivity ($\boldsymbol{k}$) and thermal resistance ($\boldsymbol{R}$) of all the building parts. The thermal resistance is; $\boldsymbol{R} = \frac{\boldsymbol{\zeta}}{\boldsymbol{k}}$. Thicknesses are examples for a BR18 building.}
  \label{tab:materials}
  \end{minipage}
\end{figure}

Based on that, the parameters are calculated from Equation \ref{eq:Rea} to \ref{eq:Cf}. 

The thermal resistances, $R\n{e,a}$, $R\n{e,i}$ and  $R\n{f,i}$;
\begin{align}
    \boldsymbol{R}\n{e,a} &= \frac{1}{U\n{windows}\cdot A\n{windows} + U\n{doors}\cdot A\n{doors} + U\n{w,a}\cdot A\n{w} + U\n{r,a}\cdot A\n{r}} \label{eq:Rea} \\ 
    &U\n{w,a} = \frac{1}{ R\n{w,s,o} + R\n{w,o} + R\n{w,insul} + \frac{R\n{w,i}}{2} } \nonumber \\
    &U\n{r,a} = \frac{1}{ \frac{R\n{r,con}}{2} + R\n{r,insul} + R\n{r,s,o}} \nonumber \\
    \boldsymbol{R}\n{e,i} &= \frac{1}{U\n{w,i}\cdot A\n{w} + U\n{r,i}\cdot A\n{r}} \label{eq:Rei}  \\ 
    &U\n{w,i} = \frac{1}{ \frac{R\n{w,i}}{2} + R\n{w,s,i}}  \nonumber \\
    &U\n{r,i} = \frac{1}{\frac{R\n{r,con}}{2} + R\n{air} + R\n{cei} + R\n{r,s,i}} \nonumber \\
    \boldsymbol{R}\n{f,i} &= \frac{1}{U\n{f,i} \cdot A\n{f}} \label{eq:Rfi}  \\ 
    &U\n{f,i} = \frac{1}{\frac{R\n{f,con}}{2} + R\n{cov} + R\n{f,s,i}}  \nonumber
\end{align}

As for the heat capacities; $C\n{e}$ is the sum of the heat capacity [kWh] in all the material on the inside of the insulation in the building envelope (concrete, air gap and ceiling). $C_f$ is the sum of the heat capacity [kWh] in the floor concrete and tiles. $C\n{i}$ is estimated from a key number of 20 $\frac{\text{kJ}}{\text{K} \cdot \text{m}^2 }$ \cite{French}, accounting for everything inside the room e.g. air and furniture. 

\begin{align}
    \boldsymbol{C}\n{e} &= \frac{1}{3.6E6} (   \zeta\n{w,i}\rho\n{w,i} C\n{w,i} \cdot A_w  \label{eq:Ce}  \\
    &+  (\zeta\n{r,air}\rho\n{r,air} C\n{r,air} 
    +  \zeta\n{r,con}\rho\n{r,con} C\n{r,con} 
    +  \zeta\n{r,cei}\rho\n{r,cei} C\n{r,cei}) \cdot A_r )  \nonumber \\ 
       \boldsymbol{C}\n{f} &= \frac{ A_f}{3.6E6} (\zeta\n{f,cov}\rho\n{f,cov} C\n{f,cov} 
    +  \zeta\n{f,con}\rho\n{f,con} C\n{f,con}) \label{eq:Cf} 
\end{align}

In Table \ref{tab:areas}, wall, floor, door and window areas are listed for the two building types along with the corresponding model parameters.

\begin{table}[htbp]
    \centering
%    \resizebox{\linewidth}{!}{%
    \begin{tabular}{>{\small}l >{\small}r >{\small}r >{\small}r}
          & &\textbf{Family house} & \textbf{Office building}\\
         $A_f $&$\text{m}^2$ & 156 & 1250\\
         $A_w $&$\text{m}^2$ & 107 & 302\\
         $A\n{doors}^* $&$\text{m}^2$ & 4 & 13 \\
         $A\n{windows}^* $&$\text{m}^2$ & 14 & 39 \\
         $\Rwa $&$\frac{\text{K}}{\text{kW}}$ & 10.398 & 2.379 \\
         $\Riw $&$\frac{\text{K}}{\text{kW}}$ & 1.190 & 0.269 \\
         $\Rfi $&$\frac{\text{K}}{\text{kW}}$ & 1.442 & 0.180 \\
         $\Cw $&$\frac{\text{kWh}}{\text{K}}$  & 7.508 & 39.527 \\
         $\Cf $&$\frac{\text{kWh}}{\text{K}}$ & 3.198 & 25.623 \\
         $\Ci $&$\frac{\text{kWh}}{\text{K}}$ & 0.876 & 6.944 
    \end{tabular} %
%}
\caption{Building dimensions and  model parameters for both buildings based on BR18. $^*$The window and door area is determined from a window-to-wall ratio of 0.11 and a door-to-wall ratio of 0.04.}
\label{tab:areas}
\end{table}

\subsection{Heat pump dimensions}
The minimum heat pump sizes are estimated from the heat loss on the coldest day. This is defined in Danish Standard DS-418 as an outdoor temperature of -12 \textdegree{}C with an indoor set point of 20 \textdegree{}C, thus $dT = 32$ \textdegree{}C.
\begin{align}
    Q\n{loss} &= (U\n{walls}\cdot A\n{walls} + U\n{windows}\cdot A\n{windows} \\
    &+ U\n{doors}\cdot A\n{doors} + U\n{roof}\cdot A\n{roof})\cdot dT
\end{align}

Note, the floor is neglected because the model assumes no heat loss through the floor. The heat losses are  2.9 $kW$ and 13.4 $kW$ for the family house and office building (danish building code of 2018) respectively. The maximum power input to the heat pump $P_{max}$ is then $\frac{f\n{cop}(40, -12)}{Q\n{loss}}$, hence 1 kW and 4.5 kW.


%=====================================
% References, variant A: internal bibliography
%=====================================
\reftitle{References}

% The following MDPI journals use author-date citation: Arts, Econometrics, Economies, Genealogy, Humanities, IJFS, JRFM, Laws, Religions, Risks, Social Sciences. For those journals, please follow the formatting guidelines on http://www.mdpi.com/authors/references
% To cite two works by the same author: \citeauthor{ref-journal-1a} (\citeyear{ref-journal-1a}, \citeyear{ref-journal-1b}). This produces: Whittaker (1967, 1975)
% To cite two works by the same author with specific pages: \citeauthor{ref-journal-3a} (\citeyear{ref-journal-3a}, p. 328; \citeyear{ref-journal-3b}, p.475). This produces: Wong (1999, p. 328; 2000, p. 475)

%=====================================
% References, variant B: external bibliography
%=====================================
%\externalbibliography{yes}
%\bibliography{your_external_BibTeX_file}

%%%%%%%%%%%%%%%%%%%%%%%%%%%%%%%%%%%%%%%%%%
%% optional
%\sampleavailability{Samples of the compounds ...... are available from the authors.}

%% for journal Sci
%\reviewreports{\\
%Reviewer 1 comments and authors’ response\\
%Reviewer 2 comments and authors’ response\\
%Reviewer 3 comments and authors’ response
%}

%%%%%%%%%%%%%%%%%%%%%%%%%%%%%%%%%%%%%%%%%%
\end{document}